%% file: main.tex
\documentclass[lettersize,journal]{IEEEtran}
\usepackage{algorithmic}
\usepackage{array}
\usepackage[caption=false,font=normalsize,labelfont=sf,textfont=sf]{subfig}
\usepackage{textcomp}
\usepackage{stfloats}
\usepackage{url}
\usepackage{verbatim}
\usepackage{graphicx}
\def\BibTeX{{\rm B\kern-.05em{\sc i\kern-.025em b}\kern-.08em
    T\kern-.1667em\lower.7ex\hbox{E}\kern-.125emX}}
\usepackage{balance}
\usepackage{times}
\usepackage{epsfig}
\usepackage{amssymb}
\usepackage{algorithm}
\usepackage{enumerate}
\usepackage{multirow}
\usepackage{multicol}
\usepackage{arydshln}
\usepackage{color}
\usepackage{amssymb}
\usepackage{amsthm}
\usepackage{booktabs}
\usepackage[accsupp]{axessibility}  
\usepackage{amsmath,amsfonts}
\usepackage{cite}

\input{defs}

\newcommand{\blue}[1] {\textcolor[rgb]{0.0,0.0,0.0}{{#1}}}  

\begin{document}
\title{Unrolling Plug-and-Play Gradient Graph Laplacian Regularizer for Image Restoration}
\author{
\IEEEauthorblockN{Jianghe Cai,  Gene Cheung, \emph{Fellow, IEEE}, Fei Chen, \emph{Member, IEEE}}
\renewcommand{\baselinestretch}{1.0}
\thanks{The work of G. Cheung was supported in part by the Natural Sciences and Engineering Research Council of Canada (NSERC) RGPIN-2019-06271, RGPAS-2019-00110. The work of F. Chen was supported in part by the National Natural Science Foundation of China (62471141). \textit{(Corresponding author: Fei Chen)}}
\thanks{J. Cai and F. Chen are with College of Computer and Data Science, Fuzhou University, Fuzhou, China (e-mail:chenfei314@fzu.edu.cn).}
\thanks{ G. Cheung is with the department of EECS, York University, 4700 Keele Street, Toronto, M3J 1P3, Canada (e-mail: genec@yorku.ca).} 
}

\markboth{ }%
{How to Use the IEEEtran \LaTeX \ Templates}

\maketitle

\maketitle
\begin{abstract}
\input{abstract.tex}
\end{abstract}
\begin{IEEEkeywords}
Image restoration, Graph signal processing, Plug-and-Play ADMM, Algorithm unrolling.
\end{IEEEkeywords}
\section{Introduction}
\label{sec:intro}
\input{intro.tex}

\section{Related Works}
\label{sec:related}
\input{related.tex}

\section{Preliminaries}
\label{sec:prelim}
\input{prelim.tex}

\section{Image Restoration via Plug-and-Play GGLR}
\label{sec:ggtv}
\input{GGTV.tex}

\section{Unrolling Plug-and-Play GGLR}
\label{sec:unroll}
\input{unroll.tex}

\section{Experiments}
\label{sec:results}
\input{results}

\section{Conclusion}
\label{sec:conclude}
\input{conclude}

\appendix
\input{append}

\bibliographystyle{IEEEbib}
\bibliography{ref}



\end{document}

%% file: defs.tex
\def\0{{\mathbf 0}}
\def\1{{\mathbf 1}}

\def\f{{\mathbf f}}

\def\n{{\mathbf n}}

\def\p{{\mathbf p}}

\def\r{{\mathbf r}}

\def\v{{\mathbf v}}

\def\x{{\mathbf x}}
\def\y{{\mathbf y}}
\def\z{{\mathbf z}}

\def\A{{\mathbf A}}

\def\D{{\mathbf D}}

\def\F{{\mathbf F}}
\def\G{{\mathbf G}}
\def\F{{\mathbf F}}
\def\H{{\mathbf H}}
\def\I{{\mathbf I}}
\def\J{{\mathbf J}}
\def\K{{\mathbf K}}
\def\L{{\mathbf L}}

\def\Q{{\mathbf Q}}

\def\V{{\mathbf V}}
\def\W{{\mathbf W}}
\def\X{{\mathbf X}}

\def\ie{{\textit{i.e.}}}
\def\eg{{\textit{e.g.}}}

\def\cE{{\mathcal E}}

\def\cG{{\mathcal G}}

\def\cL{{\mathcal L}}

\def\cN{{\mathcal N}}

\def\balpha{{\boldsymbol \alpha}}

\def\bSigma{{\boldsymbol \Sigma}}

\def\blambda{{\boldsymbol \lambda}}

%% file: abstract.tex
Generic deep learning (DL) networks for image restoration like denoising and interpolation lack mathematical interpretability, require voluminous training data to tune a large parameter set, and are fragile in the face of covariate shift. 
To address these shortcomings, we build interpretable networks by unrolling variants of a graph-based optimization algorithm of different complexities.
Specifically, for a general linear image formation model, we first formulate a convex quadratic programming (QP) problem with a new $\ell_2$-norm graph smoothness prior called gradient graph Laplacian regularizer (GGLR) that promotes piecewise planar (PWP) signal reconstruction. 
To solve the posed unconstrained QP problem, instead of computing a linear system solution straightforwardly, we introduce a variable number of auxiliary variables and correspondingly design a family of ADMM algorithms.
We then unroll them into variable-complexity feed-forward networks, amenable to parameter tuning via back-propagation. 
More complex unrolled networks require more labeled data to train more parameters, but have better overall performance. 
The unrolled networks have periodic insertions of a graph learning module, akin to a self-attention mechanism in a transformer architecture, to learn pairwise similarity structure inherent in data.
Experimental results show that our unrolled networks perform competitively to generic DL networks in image restoration quality while using only a fraction of parameters, and demonstrate improved robustness to covariate shift.

%% file: intro.tex
\IEEEPARstart{I}{mage} restoration, such as denoising and interpolation, aims to recover the original image given only a noise-corrupted / sub-sampled / filtered observation.
Existing image restoration methods can be broadly categorized into \textit{model-based} and \textit{data-driven} approaches. 
To regularize an ill-posed signal restoration problem, model-based methods adopt mathematically defined signal priors such as sparse representation \cite{sparse}, \textit{total variation} (TV) \cite{gtv2}, and low-rank matrices \cite{Gu2014WeightedNN}. 
However, while these assumed priors are easily interpretable, they tend to be overly restrictive in assumptions, resulting in sub-par performance in practical scenarios.

In contrast, advances in \textit{deep learning} (DL) have led to \textit{deep neural nets} (DNN) that achieve state-of-the-art (SOTA) performance after intensive training on large datasets \cite{DL1,DL2}. 
However, generic DNNs have known drawbacks.
First, generic DNN architectures operate like ``black boxes'' that are difficult to interpret. 
Second, generic DNNs require tuning of large parameter sets from huge training data, which may be impractical for scenarios where it is difficult or expensive to obtain large labeled datasets.
Training millions of parameters on energy-hungry server farms is also environmentally unfriendly.
Third, in the event of \textit{covariate shift}---where statistics of training and testing data differ \cite{mismatch1,mismatch2}---performance of trained DNNs can degrade precipitously. 

Instead of generic off-the-shelf DNNs, \textit{algorithm unrolling}---implementing iterations of an iterative optimization algorithm as neural layers for end-to-end parameter tuning---offers an alternative approach to build problem-specific networks \cite{Monga2021}.
Unrolled networks can be competitive in performance; specifically, \cite{Yu2023WhiteBoxTV} recently showed that multi-head self-attention operator and multi-layer perceptron in transformer can be interpreted as gradient descent and sparsifying operators that alternately minimize a sparse rate reduction (SRR) objective.
While the unrolled network in \cite{Yu2023WhiteBoxTV}---called a ``white-box'' transformer---is 100\% mathematically interpretable and on-par with SOTA transformers in a range of vision tasks, tuning of the resulting large parameter set is still formidable. 

Inspired by \cite{Yu2023WhiteBoxTV}, in this paper, leveraging recent progress in \textit{graph signal processing} (GSP) \cite{ortega18ieee,cheung18} to solve signal restoration problems \cite{pang17,liu17,liu19,bai19,dinesh20,zeng20,chen2021fast,dinesh22}, we build neural networks by unrolling variants of a graph-based optimization algorithm of different complexities. 
Like \cite{Yu2023WhiteBoxTV}, our unrolled networks are mathematically interpretable, but because our parameter set of variable size is dramatically smaller, the required training dataset is also small, resulting in networks that are more robust to covariate shift.

Specifically, given a general linear image formation model, we first formulate a convex \textit{quadratic programming} (QP) optimization problem using a recent $\ell_2$-norm graph smoothness prior \textit{gradient graph Laplacian regularizer} (GGLR) \cite{chen24tsp} that promotes \textit{piecewise planar} (PWP) signal reconstruction. 
To solve the posed unconstrained QP problem, instead of computing a linear system solution straightforwardly using a known solver such as \textit{conjugate gradient} (CG) \cite{CG1986}, we introduce a varying number of auxiliary variables corresponding to sums of GGLRs on pixel rows and columns of a target patch.
To address the modified multi-variable QP optimization, we design a family of \textit{alternating direction method of multipliers} (ADMM) algorithms \cite{Stanley_PnP_admm} and unroll them into feed-forward networks---amenable to parameter optimization via back-propagation.
We call our networks \textit{unrolled Plug-and-Play GGLR} (UPnPGGLR).
More complex UPnPGGLRs contain more parameters that required training on larger datasets, but often lead to better overall performance. 

Crucially, our UPnPGGLRs have periodic insertions of a graph learning module that learns pairwise similarity structure inherent in data. 
We show that this module is akin to a self-attention mechanism \cite{bahdanau14} in a conventional transformer architecture \cite{vaswani17attention}, but doing so in a much more parameter-efficient manner, thanks to a shallow \textit{convolutional neural net} (CNN) implementation that maps input embeddings to low-dimensional feature vectors for edge weight computation. 
This results in a lightweight transformer-like neural net that is also mathematically interpretable\footnote{\blue{Common in algorithm unrolling \cite{Monga2021},  ``interpretable" here means that each unrolled neural layer corresponds to a specific iteration of a model-based iterative optimization algorithm. In our work, we construct a graph learning module for graph-based optimization, where the edge weight learning and normalization components can be interpreted respectively as attention weight learning and the softmax operation in a self-attention mechanism.}}, similar in philosophy\footnote{\blue{While \cite{Yu2023WhiteBoxTV} designed an optimization algorithm minimizing a sparse rate reduction (SRR) objective, we design a low-pass filter minimizing GGLR, interleaving with a graph learning module. 
In our case, each graph learning module corresponds to attention weight learning, while the GGLR-based optimization corresponds to attention-based filtering.}} to \cite{Yu2023WhiteBoxTV}.

Experimental results for a range of image restoration tasks---denoising, interpolation and non-blind deblurring---show that our UPnPGGLRs perform competitively to SOTA generic DL networks, while using only a tiny fraction of parameters (\eg, fewer than $1\%$ of parameters in Restormer \cite{Zamir2021RestormerET} for denoising), and demonstrate improved robustness to covariate shift.
In summary, our main contributions are
\begin{enumerate}
\item For graph learning, we adopt shallow CNNs to learn low-dimensional features for edge weight computation.
This amounts to a parameter-efficient variant of the self-attention mechanism typical in a transformer architecture.
\item Starting from a linear image formation model, we formulate a restoration problem using $\ell_2$-norm GGLR as prior, resulting in an unconstrained QP optimization problem.
\item We introduce a varying number of auxiliary variables covering different GGLR subset sums, leading to a family of ADMM algorithms of different complexities. 
\item We unroll our family of algorithms into feed-forward networks (UPnPGGLRs) for tuning of a variable number of parameters via back-propagation.
In image restoration scenarios such as denoising, interpolation and non-blind deblurring, we show that our interpretable UPnPGGLRs achieve SOTA restoration quality while employing significantly fewer parameters. We demonstrate also improved robustness to covariate shift.
\end{enumerate}

The rest of the paper is organized as follows.
We first overview related works in Section\;\ref{sec:related}. 
We review essential graph definitions and graph smoothness priors in Section\;\ref{sec:prelim}.
We formulate the image restoration problem using GGLR as prior, and describe our family of corresponding ADMM-based optimization algorithms of variable complexities in Section\;\ref{sec:ggtv}.
We describe our unrolled network architecture, including periodic insertions of the important graph learning module in Section\;\ref{sec:unroll}.
Finally, experiments and conclusion are presented in Section\;\ref{sec:results} and \ref{sec:conclude}, respectively.

\vspace{0.05in}
\noindent
\textbf{Notation:}
Vectors and matrices are written in bold lowercase and uppercase letters, respectively.
The $(i,j)$ element and the $j$-th column of a matrix $\mathbf{A}$ are denoted by $A_{i,j}$ and $\mathbf{a}_{j}$, respectively.
The $i$-th element in the vector $\v$ is denoted by $v_{i}$.
The square identity matrix of rank $N$ is denoted by $\mathbf{I}_N$, the $M$-by-$N$ zero matrix is denoted by $\mathbf{0}_{M,N}$, and the vector of all ones / zeros of length $N$ is denoted by $\mathbf{1}_N$ / $\mathbf{0}_N$, respectively.

%% file: related.tex
\subsection{Learning-based Image Restoration}

Powerful generic DL networks are prevalent in image restoration.
For example, DnCNN \cite{DnCNN} utilizes residual learning and batch normalization to construct a DL network architecture, resulting in SOTA image denoising performance.
By taking a tunable noise level map as input, FFDNet \cite{FFDNet} can handle more complex noise scenarios.
CBDNet \cite{CBDNet} employs a noise estimation sub-net and integrates it into the network to accomplish blind restoration.
To build a CNN capable of handling several noise levels, Vemulapalli \textit{et al}. \cite{crf} employ conditional random field for regularization. Recently, transformer-based models have achieved superior results in image restoration \cite{Liang2021SwinIRIR, Zamir2021RestormerET, wang2022uformer, zamir2021multi}.
However, these models have substantially larger network sizes and require tuning of an ever-growing number of parameters to achieve such high performance. 
Moreover, they are not easy to generalize and suffer steep performance degradation when faced with mismatched statistics between training and test data \cite{Chen2023MaskedIT, Yoshida2022UnrollingGT}.
In contrast, our UPnPGGLRs achieve comparable denoising performance while employing dramatically fewer parameters and are more robust to covariate shift.

\subsection{Model-based Image Restoration}

Defining signal priors using mathematical models was widespread in early image restoration literature; popular model-based denoisers include bilateral filter \cite{bilateral}, non-local means (NLM) \cite{NLM2005}, and BM3D \cite{BM3D}. 
More recent model-based methods including trilateral weighted sparse coding (TWSC) \cite{xu18} and non-local self-similarity (NSS)\cite{hou20}.
While these methods are interpretable and robust to covariate shift, we will demonstrate in Section\;\ref{sec:results} that their denoising performances are noticeably sub-par compared to SOTA DNNs, due to their overly restrictive assumed models.
In contrast, \textit{our UPnPGGLRs achieve competitive performance while enabling interpretability and covariate shift robustness, thanks to our flexible graph model that is parameterized by low-dimensional features learned from data.}

Recently, graph smoothness priors from the GSP field \cite{ortega18ieee,cheung18}---study of discrete signals on irregular data kernels described by graphs---have been successfully applied to image restoration problems such as denoising \cite{pang17}, dequantization \cite{liu17}, and deblurring \cite{bai19}.
In particular, \textit{graph Laplacian regularizer} (GLR) \cite{pang17} is popular, due to its effective promotion of \textit{piecewise constant} (PWC) signal reconstruction and its convenient quadratic form. 
In this paper, we employ instead its generalization GGLR \cite{chen24tsp} defined on gradient graphs that promotes PWP signal reconstruction.

Note that GGLR was studied only from a model-based perspective in \cite{chen24tsp}, where the minimization of the $\ell_2$-norm objective is solved by computing a system of linear equations directly using CG. 
In contrast, in this paper we demonstrate how GGLR can be practically implemented in combination with recent deep learning technologies; in particular, for image restoration we first introduce a varying number of auxiliary variables to the $\ell_2$-norm objective, then design a family of ADMM algorithms that unroll into feed-forward networks of different complexities for data-driven parameter learning.
Moreover, we show how pairwise similarity structures can be learned from data---akin to the self-attention mechanism in transformers---in order to specify the underlying graphs and the GGLR prior.
We show in Section\;\ref{sec:results} that our unrolled networks significantly outperform solving the linear system directly as done in \cite{chen24tsp}.

Unrolling of GLR for image denoising was proposed in \cite{deepglr}, and unrolling of \textit{graph total variation} (GTV) was done in \cite{deepgtv}.
Unlike \cite{deepglr}, we unroll GGLR instead using a varying number of auxiliary variables, resulting in a family of PnP-GGLR algorithms  \cite{Stanley_PnP_admm} and corresponding variable-complexity feed-forward networks with different numbers of tunable parameters.
We show in Section\;\ref{sec:results} that unrolling of GGLR outperforms unrolling of GLR and GTV in our experiments.

\subsection{\blue{Plug-and-Play Methods for Image Restoration}}

\blue{
Beyond the early PnP-ADMM work \cite{Stanley_PnP_admm}, there has been a plethora of PnP methods for image restoration; recent ones include \cite{EqDRUNtet,SNORE,Ebner2024,Reddy2024}.
\cite{EqDRUNtet} enforced equivariance to certain transformations, such as rotations and reflections, on the denoiser, resulting in improvement in stability and reconstruction of the restoration algorithm.
\cite{SNORE}, based on stochastic regularization, applied the denoiser only
on images with noise of the adequate level, leading to a stochastic gradient descent algorithm that solves the ill-posed inverse problem.
\cite{Ebner2024} extended the PnP idea by studying families of PnP iterations, where each iteration has an accompanied denoiser.
\cite{Reddy2024} proposed an optimization framework based on proximal gradient, where the gradient update ensures the iterates remain in the solution-space, and an off-the-shelf denoiser can replace the proximal operator.
}

\blue{
What is common among these PnP methods is that an implicit pre-trained denoiser is used as a basic building block in a larger optimization framework.
In contrast, we employ a new GGLR regularization leading to a family of ADMM algorithms, that unroll into feed-forward networks for data-driven parameter tuning.
That means unlike \cite{EqDRUNtet,SNORE,Ebner2024,Reddy2024}, our ``denoiser" is specific and interpretable. 
Further, none of \cite{EqDRUNtet,SNORE,Ebner2024,Reddy2024} explicitly contain a graph learning module that constitutes a self-attention mechanism---a key contribution in our work.
Nonetheless, to demonstrate the superiority of our method, we compared UPnPGGLR against \cite{EqDRUNtet} and \cite{SNORE} on image restoration tasks in our experiments.
}

\subsection{Deep Algorithm Unrolling}

The seminal work in algorithm unrolling \cite{Monga2021} is  the unrolling of \textit{iterative soft-thresholding algorithm} (ISTA) in sparse coding into \textit{Learned} ISTA (LISTA) \cite{gregor10}.
Since then, algorithm unrolling has been employed in a wide variety of algorithms for different applications, such as non-negative matrix factorization for speech processing \cite{Hershey2014DeepUM}, proximal gradient descent for medical imaging \cite{Yang2020ADMMCSNetAD}. 
Instead of unrolling a \textit{fixed} iterative algorithm, in this paper we vary the number of auxiliary variables to induce a family of PnP-GGLR algorithms, resulting in variable-complexity feed-forward networks for parameter tuning.

\blue{
We note that the original PnP-ADMM \cite{Stanley_PnP_admm} and many subsequent PnP works \cite{EqDRUNtet,SNORE,Ebner2024,Reddy2024} assume an implicit pre-trained denoiser for solving a denoising sub-problem in an overall optimization framework.
In contrast, our proposed family of algorithms focuses exclusively on GGLR as the chosen signal prior for image restoration.
Nonetheless, we liberally abuse the terminology ``PnP" here to mean that our setup permits the flexible use of any linear image formation model (accommodating applications such as denoising, interpolation and non-blind deblurring) and the introduction of a varying number of auxiliary variables, leading to unrolled networks of variable complexity.
}

%% file: prelim.tex
\subsection{GSP Definitions}
\label{subsec:defn}

A graph $\cG(\cN,\cE,\W)$ is defined by a node set $\cN = \{1, \ldots, N\}$ and an edge set $\cE$, where $(i,j) \in \cE$ means nodes $i,j \in \cN$ are connected with weight $w_{i,j} = W_{i,j} \in \mathbb{R}$.
Fig.\;\ref{fig:graphEx}(a) shows an example of a 3-node line graph.
We assume edges are undirected, and thus \textit{adjacency matrix} $\W \in \mathbb{R}^{N \times N}$ is symmetric. 
The \textit{combinatorial graph Laplacian matrix} is defined as $\L \triangleq \text{diag}(\W \1_N) - \W \in \mathbb{R}^{N \times N}$, where $\text{diag}(\v)$ returns a diagonal matrix with $\v$ along its diagonal.
Real and symmetric $\L$ is \textit{positive semi-definite} (PSD) if $w_{i,j} \geq 0, \forall i,j$, \ie, $\x^\top \L \x \geq 0, \forall \x$ \cite{cheung18}. 

Real and symmetric $\L$ can be eigen-decomposed to $\L = \V \bSigma \V^\top$, where $\V$ contains eigenvectors of $\L$ as columns, and $\bSigma = \text{diag}(\lambda_1, \ldots, \lambda_N)$ is a diagonal matrix with real and non-negative eigenvalues $0 = \lambda_1 \leq \lambda_2 \leq \ldots \leq \lambda_N$ along its diagonal.
Eigen-pair $(\lambda_k,\v_k)$ are the $k$-th graph frequency and Fourier mode for $\cG$, respectively.
$\tilde{\x} = \V^\top \x$ is the \textit{graph Fourier transform} (GFT) of signal $\x$ \cite{ortega18ieee}.

\subsection{Graph Laplacian Regularizer}

To regularize an ill-posed graph signal restoration problem, the \textit{graph Laplacian regularizer} (GLR) is often used due to its convenient quadratic form \cite{pang17}. 
GLR for a signal $\x \in \mathbb{R}^N$ is
\begin{align}
\x^\top \L \x &= \sum_{(i,j) \in \cE} w_{i,j} (x_i - x_j)^2 = \sum_k \lambda_k \tilde{x}_k^2
\label{eq:GLR} 
\end{align}
where $\tilde{x}_k = \v_k^\top \x$ is the $k$-th graph frequency coefficient for signal $\x$.
Thus, a \textit{low-pass} signal with energies $\tilde{x}_k$'s concentrated in low graph frequencies $\lambda_k$'s would induce a small GLR value in \eqref{eq:GLR}. 

Edge weights $w_{i,j}$ can be defined in a \textit{signal-dependent} manner, specifically,
\begin{align}
w_{i,j}(x_i,x_j) = \exp \left(- \frac{\|\f_i - \f_j\|^2_2}{\sigma_f^2} - \frac{|x_i - x_j|^2}{\sigma_x^2} \right) ,
\label{eq:GLRweight}
\end{align}
where $\f_i \in \mathbb{R}^K$ is a \textit{feature vector} for node $i$, and $\sigma_f$ and $\sigma_x$ are parameters.
\eqref{eq:GLRweight} is signal-dependent because of its dependency on signal samples $x_i$ and $x_j$.
Using edge weight definition \eqref{eq:GLRweight} means GLR is also signal-dependent. 
Iteratively applying signal-dependent GLR promotes \textit{piecewise constant} (PWC) signal reconstruction; see Appendix\;\ref{append:GLR} for a detailed discussion.

\subsection{Gradient Graph Laplacian Regularizer}
\label{subsec:GGLR}

\begin{figure}[t]
\begin{center}
\includegraphics[width=0.75\linewidth]{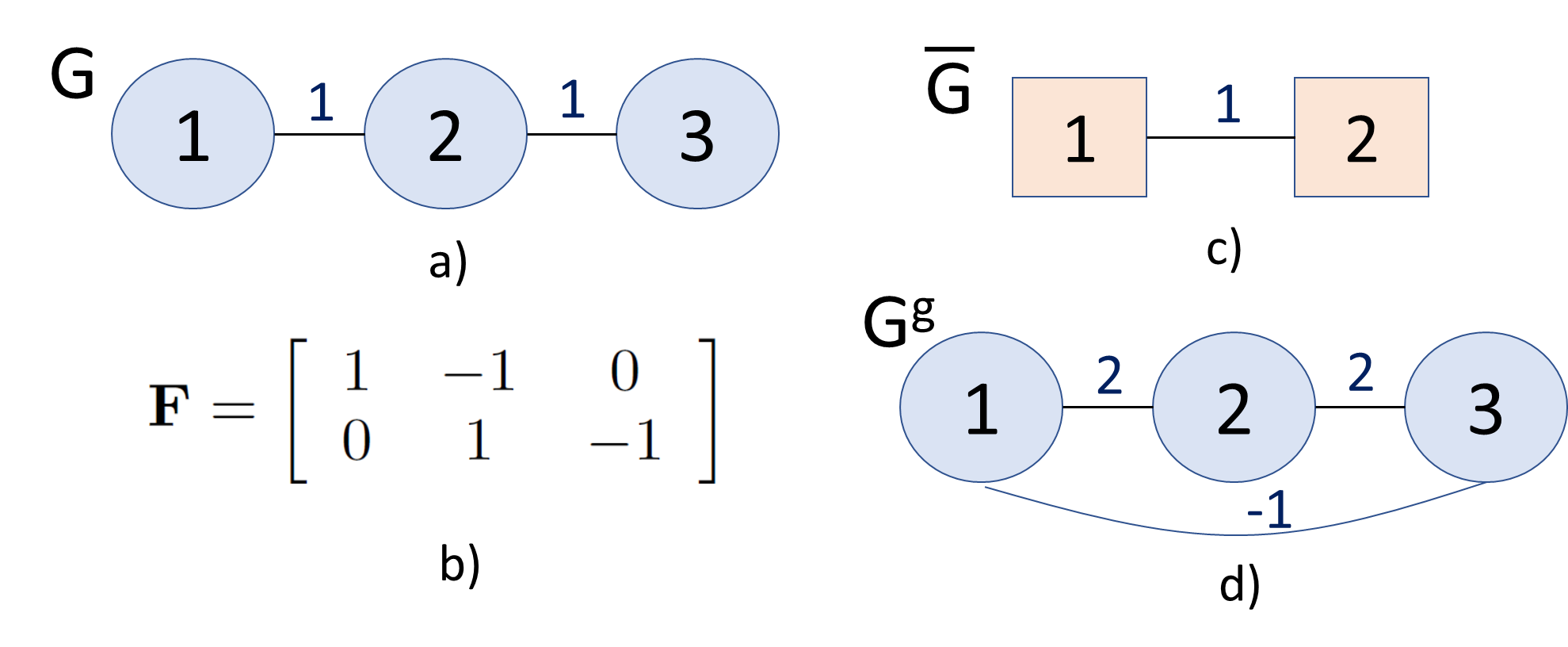}
\end{center}
\vspace{-0.3in}
\caption{A 3-node line graph $\cG$ in (a), a gradient operator $\F \in \mathbb{R}^{2 \times 3}$ for a row or column of 3 pixels in (b), gradient graph $\bar{\cG}$ in (c), and resulting GNG $\cG^g$ in (d)---a signed graph with both positive and negative edges \cite{chen2021fast}. } 
\vspace{-0.15in}
\label{fig:graphEx}
\end{figure}

GGLR \cite{chen24tsp} applies GLR to the gradient of signal $\x$ instead of $\x$ directly; for 1D signals, a (piecewise) constant signal gradient means a \textit{(piecewise) linear} (PWL) signal. 
For images, GGLR can mean applying GLR to horizontal/vertical image gradients separately.
To compute horizontal/vertical gradient $\balpha \in \mathbb{R}^{N-1}$ for a pixel row/column $\x \in \mathbb{R}^N$, we first define a \textit{gradient operator} $\F \in \mathbb{R}^{N-1 \times N}$:
\begin{align}
F_{i,j} &= \left\{ \begin{array}{ll}
1 & \mbox{if}~ i=j \\
-1 & \mbox{if}~i=j-1 \\
0 & \mbox{o.w.}
\end{array} \right. .
\label{eq:gradientOp}
\end{align}
Note that $\F \1 = \0$ and $\F$ is full row-rank \cite{chen24tsp}. 
We then compute horizontal/vertical gradients $\balpha \in \mathbb{R}^{N-1}$ as
\begin{align}
\balpha = \F \x .
\label{eq:balpha}
\end{align}
A positive \textit{gradient graph} $\bar{\cG}$ is constructed to connect gradients $i$ and $j$ with non-negative weight $\bar{w}_{i,j}$, computed in a \textit{gradient-dependent} manner, similarly to \eqref{eq:GLRweight}, as
\begin{align}
\bar{w}_{i,j}(\alpha_i,\alpha_j) = \exp \left( - \frac{\|\f_i - \f_j\|^2_2}{\sigma_f^2} - \frac{|\alpha_i - \alpha_j|^2}{\sigma_a^2} \right) ,
\label{eq:gradEdgeWeight}
\end{align}
where $\f_i \in \mathbb{R}^K$ is a feature vector for gradient $i$, and \blue{$\sigma_f$ and $\sigma_a$ are learnable parameters.}
See Fig.\;\ref{fig:graphEx}(b) and (c) for an example of gradient operator $\F$ for a 3-pixel row and the corresponding gradient graph $\bar{\cG}$ for $\balpha \in \mathbb{R}^2$, respectively.

Define a \textit{gradient adjacency matrix} $\bar{\W} \in \mathbb{R}^{(N-1) \times (N-1)}$ using the set of edge weights $\{\bar{w}_{i,j}\}$. 
Correspondingly, define \textit{gradient graph Laplacian} matrix as $\bar{\L} \triangleq \text{diag}(\bar{\W} \1_N) - \bar{\W}$. 
Finally, GGLR is defined as
\begin{align}
\balpha^\top \bar{\L} \balpha = \x^\top \underbrace{\F^\top \bar{\L} \F}_{\cL} \x 
= \x^\top \cL \x 
\label{eq:GGLR}
\end{align}
where $\cL$ is a \textit{gradient-induced nodal graph} (GNG) Laplacian for a GNG $\cG^g$.
\cite{chen24tsp} proved that the eigen-subspace corresponding to first eigenvalue $\lambda_1 = 0$ of $\cL$ has dimension $2$ (\ie, multiplicity of eigenvalue $\lambda_1$ is two), and spans the space of all linear signals $\x = m\, i + c$, where $i$ is the node index of line graph $\cG$, and $m$ and $c$ are the respective slope and $y$-intercept of the line.
Thus, using GGLR $\x^\top \cL \x$ as signal prior ``promotes'' linear signal reconstruction (\ie, $\x^\top \cL \x = 0$ for any linear $\x$), and by extension signal-dependent GGLR promotes piecewise linear signal reconstruction.
See Appendix\;\ref{append:GGLR} for a detailed discussion. 

Fig.\;\ref{fig:graphEx}(d) shows an example of a GNG $\cG^g$ corresponding to the gradient graph $\bar{\cG}$ in (c).
Note that in general, $\cG^g$ is a \textit{signed} graph with both positive and negative edges; 
a signed graph is necessary so that GGLR $\x^\top \cL \x = 0$ for any linear signal $\x$, as earlier discussed.
However, given $\bar{\L}$ is PSD, $\cL = \F^\top \bar{\L} \F $ must also be PSD, though $\cG^g$ is a signed graph.

%% file: GGTV.tex
\subsection{GGLR for Image Patches}

\subsubsection{Horizontal/Vertical Line Graphs for Horizontal/Vertical Gradients}
Assuming a target patch is PWP, its pixel rows and columns must be PWL. 
Thus, to restore an $N$-by-$N$ image patch, we first employ GGLR as a prior for $N$ individual pixel rows and $N$ columns in the patch.
Denote by $\x \in \mathbb{R}^{N^2}$ the vectorized version of the patch by scanning pixels row-by-row.
Denote by $\H_k, \G_k \in \{0,1\}^{N \times N^2}$ the respective sampling matrix that picks out $N$ pixels of the $k$-th row/column from $N^2$ entries in $\x$, \ie, 
\begin{align}
H_{k,i,j} = \left\{ \begin{array}{ll} 
1 & \mbox{if entry $j$ is the $i$-th pixel in $k$-th row} \\
0 & \mbox{o.w.}
\end{array} \right. 
\\
G_{k,i,j} = \left\{ \begin{array}{ll} 
1 & \mbox{if entry $j$ is the $i$-th pixel in $k$-th column} \\
0 & \mbox{o.w.}
\end{array} \right. 
\end{align}
For example, $\H_1$ and $\G_2$ that pick out the first row and the second column of a $2 \times 2$ pixel patch are
\begin{align}
\H_1 = \left[ \begin{array}{cccc}
1 & 0 & 0 & 0 \\
0 & 1 & 0 & 0
\end{array} \right], ~~
\G_2 = \left[ \begin{array}{cccc}
0 & 1 & 0 & 0 \\
0 & 0 & 0 & 1
\end{array} \right] .
\end{align}

For each $k$-th row/column of $N$ pixels, we compute horizontal/vertical gradient $\balpha$ using \eqref{eq:balpha}.
Then, we connect gradients $i$ and $j$ using a positive edge with weight $\bar{w}_{i,j}$ defined in \eqref{eq:gradEdgeWeight}. 
The resulting gradient line graph $\bar{\cG}^{h,r}_k$ or $\bar{\cG}^{v,c}_k$ has corresponding graph Laplacian matrix $\bar{\L}^{h,r}_k$ or $\bar{\L}^{v,c}_k$, which is used to define GNG Laplacian $\cL^{h,r}_k \triangleq \F^\top \bar{\L}^{h,r}_k \F$ or $\cL^{v,c}_k \triangleq \F^\top \bar{\L}^{v,c}_k \F$.
Finally, the GNG Laplacians are used to define GGLR:
$\x^\top \H_k^\top \cL^{h,r}_k \H_k \x$ or $\x^\top \G_k^\top \cL^{v,c}_k \G_k \x$.

\vspace{0.05in}
\subsubsection{Vertical/Horizontal Line Graphs for Horizontal/Vertical Gradients}
While the previous GNG Laplacians can promote PWL pixel row and column reconstruction, it is not sufficient to ensure PWP signal reconstruction. 
For example, consider the following $3 \times 3$ image patch with linear pixel rows and columns, and yet the patch is not planar:
\begin{align}
\X = \left[ \begin{array}{ccc}
1 & 0 & -1 \\
0 & 0 & 0 \\
-1 & 0 & 1 
\end{array} \right]. 
\end{align}
Towards PWP signal reconstruction, we construct also vertical/horizontal line graphs to connect horizontal/vertical gradients as well.
Specifically, denote by $\J_k, \K_k \in \{0,1\}^{2N \times N^2}$ the respective sampling matrix that picks out $2N$ pixels of the $k$-th and $(k+1)$-th columns/rows from $N^2$ entries in $\x$, \ie,

\vspace{-0.1in}
\begin{small}
\begin{align}
J_{k,i,j} = \left\{ \begin{array}{ll} 
1 & \mbox{if $i$ is odd and $j$ is $\lceil \frac{i}{2} \rceil$-th pixel in $k$-th col} \\
1 & \mbox{if $i$ is even and $j$ is $ \frac{i}{2}$-th pixel in $(k+1)$-th col} \\
0 & \mbox{o.w.}
\end{array} \right. 
\\
K_{k,i,j} = \left\{ \begin{array}{ll} 
1 & \mbox{if $i$ is odd and $j$ is $\lceil \frac{i}{2} \rceil$-th pixel in $k$-th row} \\
1 & \mbox{if $i$ is even and $j$ is $\frac{i}{2}$-th pixel in $(k+1)$-th row} \\
0 & \mbox{o.w.}
\end{array} \right. 
\end{align}
\end{small}
For example, $\J_1$ that picks out the 1st and 2nd columns of a $3 \times 3$ pixel patch is
\begin{align}
\J_1 &= \left[ \begin{array}{ccccccccc}
1 & 0 & 0 & 0 & 0 & 0 & 0 & 0 & 0 \\
0 & 1 & 0 & 0 & 0 & 0 & 0 & 0 & 0 \\
0 & 0 & 0 & 1 & 0 & 0 & 0 & 0 & 0 \\
0 & 0 & 0 & 0 & 1 & 0 & 0 & 0 & 0 \\
0 & 0 & 0 & 0 & 0 & 0 & 1 & 0 & 0 \\
0 & 0 & 0 & 0 & 0 & 0 & 0 & 1 & 0 \\
\end{array} \right].
\end{align}
We correspondingly define a new gradient operator $\tilde{\F} \in \mathbb{R}^{N \times 2N}$ as a counterpart to $\F$ in \eqref{eq:gradientOp} as
\begin{align}
\tilde{F}_{i,j} &= \left\{ \begin{array}{ll}
1 & \mbox{if $j$ is odd and $i= \lceil \frac{j}{2} \rceil$} \\
-1 & \mbox{if $j$ is event and $i= \frac{j}{2}$} \\
0 & \mbox{o.w.}
\end{array} \right. .
\label{eq:gradientOp2}
\end{align}
For example, for a $3 \times 3$ patch, $\tilde{\F}$ is
\begin{align}
\tilde{\F} = \left[ \begin{array}{cccccc}
1 & -1 & 0 & 0 & 0 & 0 \\
0 & 0 & 1 & -1 & 0 & 0 \\
0 & 0 & 0 & 0 & 1 & -1
\end{array} \right] .
\end{align}

We can now construct vertical/horizontal line graphs $\bar{\cG}^{v,r}_k$ or $\bar{\cG}^{h,c}_k$ to connect horizontal/vertical gradients $\tilde{\F} \J_k \x$ or $\tilde{\F} \K_k \x$ with weight $\bar{w}_{i,j}$ defined in \eqref{eq:gradEdgeWeight}. 
The resulting gradient graph Laplacian matrices are $\bar{\L}^{v,r}_k$ and $\bar{\L}^{h,c}_k$, which are used to define GNG Laplacian $\cL^{v,r}_k \triangleq \tilde{\F}^\top \bar{\L}^{v,r}_k \tilde{\F}$ or $\cL^{h,c}_k \triangleq \tilde{\F}^\top \bar{\L}^{h,c}_k \tilde{\F}$.
Finally, the GNG Laplacians are used to define GGLR:
$\x^\top \J_k^\top \cL^{v,r}_k \J_k \x$ or $\x^\top \K_k^\top \cL^{h,c}_k \K_k \x$.

\subsubsection{Linear Image Formation Model}
Consider a standard linear image formation for observation $\y \in \mathbb{R}^M$ from ground truth signal $\x \in \mathbb{R}^{N^2}$:
\begin{align}
\y = \A \x + \n
\label{eq:formation}
\end{align}
where $\A \in \mathbb{R}^{M \times {N^2}}$ is a degradation matrix, and $\n \in \mathbb{R}^M$ is an additive noise.
Different $\A$'s imply different restoration tasks: if $\mathbf{A}$ is an identity matrix, then optimization \eqref{eq:obj_restore} is a denoising problem; if $\A \in \{0,1\}^{M \times {N^2}}$ is a sampling matrix, then \eqref{eq:obj_restore} is an interpolation problem; if $\A \in \mathbb{R}^{N^2 \times N^2}$ is a low-pass blur filter, then \eqref{eq:obj_restore} is a non-blind deblurring problem.

\subsubsection{Optimization Objective} 
Given image formation model \eqref{eq:formation} and assuming noise $\n$ is Gaussian, the optimization using GGLR as prior is
\begin{align}
\min_{\x} ~& \|\y - \A \x \|^2_2 + \mu \, \x^\top \underbrace{\left( \sum_{k=1}^N \H_k ^\top \cL^{h,r}_k \H_k + \G_k ^\top \cL^{v,c}_k \G_k \right)}_{\cL} \x 
\nonumber \\
& + \tilde{\mu} \, \x^\top \underbrace{\left( \sum_{k=1}^{N-1} \J_k ^\top \cL^{v,r}_k \J_k + \K_k ^\top \cL^{h,c}_k \K_k \right)}_{\tilde{\cL}} \x 
\label{eq:obj_restore}
\end{align}
where $\mu, \tilde{\mu} > 0$ are weight parameters.
Importantly, note that the priors $\x^\top \cL \x$ and $\x^\top \tilde{\cL} \x$ are sums of GGLR terms; we leverage this fact when introducing a variable number of auxiliary variables in the objective in the sequel.

\subsection{Plug-and-Play GGLR}

Given that objective \eqref{eq:obj_restore} is convex ($\cL$ and $\tilde{\cL}$ are provably PSD) and differentiable w.r.t. variable $\x$, one can compute solution $\x^*$ by solving the linear system:
\begin{align}
\left(\A^\top \A + \mu \cL + \tilde{\mu} \tilde{\cL} \right)\x^* = \A^\top \y .
\label{eq:linSys}
\end{align}
See \cite{chen24tsp} for sufficient conditions when $\A^\top \A + \mu \cL + \tilde{\mu} \tilde{\cL}$ is positive definite (PD) and thus invertible.
One can compute $\x^*$ in \eqref{eq:linSys} efficiently using \textit{conjugate gradient} (CG) without matrix inverse \cite{conjugate_grad}---complexity of CG is linear assuming coefficient matrix $\A^\top \A + \mu \cL + \tilde{\mu} \tilde{\cL}$ is sparse, symmetric and PD.
Instead, to induce a richer set of network parameters for end-to-end optimization after unrolling (to be discussed in Section\;\ref{sec:unrolling}), we intentionally introduce auxiliary variables to the objective \eqref{eq:obj_restore} before solving the multi-variable optimization via an ADMM algorithm \cite{Stanley_PnP_admm}.

\subsubsection{ADMM (single auxiliary variable)}
We first introduce \textit{one} auxiliary variable $\z$ and constraint $\x = \z$, so that \eqref{eq:obj_restore} can be rewritten using the augmented Lagrangian method as

\vspace{-0.05in}
\begin{small}
\begin{align}
\min_{\x,\z}  \|\y - \A \x\|^2_2 + \z^\top (\mu \cL + \tilde{\mu} \tilde{\cL}) \z + \blambda^\top (\x - \z) + \frac{\rho}{2} \left\|\x - \z \right\|^2_2 
\label{eq:obj_ADMM}
\end{align}
\end{small}\noindent 
where $\blambda \in \mathbb{R}^N$ is a Lagrange multiplier vector, and $\rho > 0$ is a weight parameter. 
To optimize \eqref{eq:obj_ADMM}, we minimize variables $\x$ and $\z$ alternately and update $\blambda$ until solution $(\x, \z, \blambda)$ converge; \ie, we solve the following sub-problems at iteration $k$:
\begin{align}
\x^{(k+1)} &= \arg \min_\x ~ \left\| \y -\A \x \right\|_{2}^{2} + \frac{\rho}{2} \| \x-\hat{\x}^{(k)} \|^2_2 
\label{eq:ADMM1} \\
\z^{(k+1)} &= \arg \min_\z \z^\top (\mu \cL + \tilde{\mu} \tilde{\cL}) \z + \frac{\rho}{2} \|\z - \hat{\z}^{(k)} \|^2_2
\label{eq:ADMM2} \\
\hat{\blambda}^{(k+1)} &= \hat{\blambda}^{(k)} + \left(\x^{(k+1)} - \z^{(k+1)} \right)
\label{eq:ADMM3}
\end{align}
where 
\begin{align}
\hat{\blambda}^{(k)} = (1/\rho) \blambda^{(k)},
\hat{\x}^{(k)} = \z^{(k)} - \hat{\blambda}^{(k)},
\hat{\z}^{(k)} = \x^{(k+1)} + \hat{\blambda}^{(k)} .
\end{align}
\eqref{eq:ADMM1} is a least-square problem independent of signal priors.
\eqref{eq:ADMM3} is a term-by-term update equation.
\eqref{eq:ADMM2} is a denoising problem for $\z$ using GGLR as prior.
Because we adopt GGLR as prior for the denoising problem in \eqref{eq:ADMM2}, we call the procedure to iteratively solve \eqref{eq:ADMM1}, \eqref{eq:ADMM2} and \eqref{eq:ADMM3} \textbf{PnP GGLR}.

\subsubsection{Solving Linear Systems via CG}
\label{subsubsec:LG}

We use CG \cite{conjugate_grad} to solve linear systems for $\x^{(k+1)}$ in \eqref{eq:ADMM1} and $\z^{(k+1)}$ in \eqref{eq:ADMM2}:
\begin{align}
(2\A^\top \A + \rho \I_N) \x^{(k+1)} &= 2\A^\top \y + \rho \, \hat{\x}^{(k)} 
\label{eq:linSys1} \\
(\I_N + \frac{2\mu}{\rho} \cL + \frac{2\tilde{\mu}}{\rho} \tilde{\cL}) \z^{(k+1)} &= \hat{\z}^{(k)} .
\label{eq:linSys2}
\end{align}
Focusing on \eqref{eq:linSys1}, we solve for $\x^{(k+1)}$ iteratively. Specifically, for each iteration $t$, we update solution $\x_{t+1}$, residual $\r_{t+1}$ and search direction $\p_{t+1}$ as
\begin{align}
\x_{t+1} &= \x_t + \alpha_t \p_t 
\label{eq:CG1} \\
\r_{t+1} &= \r_t - \alpha_t (2 \A^\top \A + \rho \I_N) \p_t 
\label{eq:CG2} \\
\p_{t+1} &= \r_{t+1} + \beta_t \p_t
\label{eq:CG3}
\end{align}
where \textit{step size} $\alpha_t$ and \textit{momentum} $\beta_t$ are computed as
\begin{align}
\alpha_t = \frac{\r_t^\top \r_t}{\p_t^\top (2 \A^\top \A + \rho \I_N) \p_t}, 
~~~~~~
\beta_t = \frac{\r_{t+1}^\top \r_{t+1}}{\r_t^\top \r_t} .
\end{align}
Solving for $\z^{(k+1)}$ in \eqref{eq:linSys2} can be done using CG similarly.

\subsubsection{ADMM (multiple auxiliary variables)}
We generalize the above ADMM optimization framework to the case where multiple auxiliary variables are introduced. 
Define $\z$ and $\tilde{\z}$ as auxiliary variables for the two GGLR summations in \eqref{eq:obj_restore}.
The augmented Lagrangian optimization objective \eqref{eq:obj_ADMM} thus becomes
\begin{align}
\min_{\x,\z,\tilde{\z}}  & ~\|\y - \A \x\|^2_2 + \mu \,  \z^\top \cL \z + \blambda^\top (\x - \z) + \frac{\rho}{2} \left\|\x - \z \right\|^2_2  
\nonumber \\
& + \tilde{\mu} \,  \tilde{\z}^\top \tilde{\cL} \tilde{\z} + \tilde{\blambda}^\top (\x - \tilde{\z}) + \frac{\tilde{\rho}}{2} \left\|\x - \tilde{\z} \right\|^2_2 
\label{eq:obj_ADMM_a2}
\end{align}
with three optimization variables. 
Similar optimization sub-routines as \eqref{eq:ADMM1}, \eqref{eq:ADMM2} and \eqref{eq:ADMM3} can be written to compute $\x^{(k+1)}$, $\z^{(k+1)}$, $\tilde{\z}^{(k+1)}$ and update $\blambda^{(k+1)}$, $\tilde{\blambda}^{(k+1)}$ in order in iteration $k$. 
See Appendix\;\ref{append:twoAux} for details.

We can generalize further and define auxiliary variables $\z_r$, $\z_c$, $\tilde{\z}_r$ and $\tilde{\z}_c$ corresponding to GGLR partial sums in \eqref{eq:obj_restore}: $\H_k^\top \cL^{h,r}_k \H_k$'s, $\G_k^\top \cL^{v,c}_k \G_k$'s, $\J_k^\top \cL^{v,r}_k \J_k$'s and $\K_k^\top \cL^{h,c}_k \K_k$'s.
This again leads to an iterative ADMM optimization procedure similar to one discussed earlier.

%% file: unroll.tex
\label{sec:unrolling}

\input{image/model.tex}

\subsection{Unrolled Network Architecture}

For a given ADMM algorithm with one or more auxiliary variable $\z$ corresponding to GGLR subset sums, we unroll it into $K$ GGLR layers as shown in Fig.\;\ref{fig:model}. 
Each GGLR layer has the same structure and parameters. 
\blue{Specifically, each layer contains a graph learning module and a GGLR-based optimization module.
For the single auxiliary variable case, a GGLR-based optimization module contains the $k$-th iteration of \eqref{eq:linSys1} and \eqref{eq:linSys2}.}
To learn edge weights from data to define GNG Laplacians, we adopt a shallow CNN$_f$ network to learn low-dimensional feature vectors $\f_i \in \mathbb{R}^K$ per pixel $i$, and then compute horizontal and vertical gradients to establish edge weights in \eqref{eq:gradEdgeWeight}.
\blue{The use of a shallow CNN for low-dimensional feature learning towards the construction of a similarity graph contributes to network parameter reduction.}
\blue{Further, in each GGLR-based optimization module, we employ a CNN$_y$ network to pre-filter the image $\mathbf{z}^{(k)}$, following the approach in \cite{deepglr}.}
Parameters $\rho$, $\mu$ and $\tilde{\mu}$ in \eqref{eq:linSys1} and \eqref{eq:linSys2} are trainable parameters in the unrolled network.

We unroll the iterative CG algorithm for solving linear system \eqref{eq:linSys1} into $L$ neural layers, where $\alpha_\mathbf{x}=\{\alpha_0, \cdots, \alpha_{L-1}\}$ and $\beta_\mathbf{x}=\{\beta_0, \cdots, \beta_{L-2}\}$ are CG parameters learned from data. 
Similarly, parameter sets $\alpha_\mathbf{z}$ and $\beta_\mathbf{z}$ are separate CG parameters learned for solving linear system \eqref{eq:linSys2}.
\blue{As shown in Fig.\;\ref{fig:model}(c), each unrolled CGD layer includes $L$ iterations of \eqref{eq:CG1} to \eqref{eq:CG3}. 
The blue lines denote computation of $\x$, the black lines denote computation of $\p$, and the green lines denote computation of $\r$.}

Similarly done in \cite{deepgtv}, the input to the algorithm is a degraded observed image 
$\mathbf{z}^{(0)}=\mathbf{y}$ and $\hat{\mathbf{\blambda}}^{(0)}=\mathbf{0}$.
After the unrolled $K$ layers, we obtain the reconstruction signal $\mathbf{x}^{K}$. 
The $\mathrm{GGLR Layer}_k$ uses the signal after $k-1$ iterations $\mathbf{z}^{(k-1)}$, $\hat{\mathbf{\blambda}}^{(k-1)}$ and a set of learned parameters 
\blue{
$\mathbf{\Theta}^{k}=\{\sigma_{a}^{(k)}, \sigma_{f}^{(k)}, \rho^{(k)},\mu^{(k)}, \tilde{\mu}^{(k)},  
\mathbf{\theta}_{f}^{(k)}, \mathbf{\theta}_{y}^{(k)}, \mathbf{\alpha}_\mathbf{x}^{(k)},\mathbf{\beta}_\mathbf{x}^{(k)}, \mathbf{\alpha}_\mathbf{z}^{(k)},\mathbf{\beta}_\mathbf{z}^{(k)}\}$ as input, where $\mathbf{\theta}_f^{(k)}$ and $\mathbf{\theta}_y^{(k)}$ denote the parameter sets of $\mathrm{CNN}_{f}$ and $\mathrm{CNN}_{y}$, respectively.}
   
To effectively train the proposed unrolling PnP GGLR (UPnPGGLR), we evaluate the difference between the recovered image $\mathbf{x}^K$ and the ground-truth image $\mathbf{x}^{gt}$ after the last GGLR layer. 
Specifically, we compute the partial derivative of the \textit{mean square error} (MSE) between $\mathbf{x}^{gt}$ and $\mathbf{x}^K$ and back-propagate it to update parameters $\{\mathbf{\Theta}^1, \cdots, \mathbf{\Theta}^K\}$ for $T$ training patches:
\begin{equation}
\mathcal{L}_{\mathbf{MSE}}(\mathbf{\Theta}^1, \cdots, \mathbf{\Theta}^K)=\frac{1}{T}\sum_{i=1}^{T} \|\x_i^K - \x_i^{gt} \|^2_2.
\label{equation9}
\end{equation}

\subsection{Self-Attention Operator in Transformer}

We argue that the periodically inserted graph learning module in our unrolled network is akin to the self-attention mechanism \cite{bahdanau14} in a conventional transformer architecture \cite{vaswani17attention}.
We first review the typical self-attention operator, defined using a linearly-transformed dot product.
Denote by $\x_i \in \mathbb{R}^B$ an input \textit{embedding} for token $i$ of $N$ total input tokens.
The \textit{affinity} $e(i,j)$ between tokens $i$ and $j$---a measure of how relevant key token $i$ is to query token $j$---is defined as the dot product between linearly-transformed $\K \x_i$ and $\Q \x_j$: 
\begin{align}
e(i,j) = (\Q \x_j)^\top (\K \x_i) .
\label{eq:SA_coeff}
\end{align}
where $\Q, \K \in \mathbb{R}^{B \times B}$ are respectively the \textit{query} and \textit{key} matrices---dense matrices learned from data.

Commonly, using a non-linear function called \textit{softmax}, \textit{attention weight} $a_{i,j}$ is computed as 
\begin{align}
a_{i,j} = \frac{\exp (e(i,j))}{\sum_{l=1}^N \exp (e(i,l))}.
\label{eq:SA_coeff2}
\end{align}
Besides converting an input set of real values $e(i,j)$'s to positive values $a_{i,j}$'s, softmax also normalizes $a_{i,j}$'s so that they sum to $1$. 

Using weights $a_{i,j}$'s, we compute the output embedding $\y_i$ for token $i$ using a \textit{value} matrix $\V \in \mathbb{R}^{B \times B}$, another dense matrix learned from data, as
\begin{align}
\y_i = \sum_{l=1}^N a_{i,l} \x_l \V .
\label{eq:SA_operator}
\end{align}
The term ``self-attention'' conveys the notion that output embeddings are computed using input embeddings.
A sequence of embedding-to-embedding mappings composes a transformer, defined by the corresponding learned matrices, $\Q$, $\K$ and $\V$ in each mapping.
Note that one can extend the framework to \textit{multi-head} attention, when we use multiple query and key matrices $\Q^{(m)}$ and $\K^{(m)}$ to compute different weights $a^{(m)}_{i,j}$'s for the same input embeddings $\x_i$ and $\x_j$.
In this case, we compute output $\y_i$ as the average of these multi-head attention weights $a^{(m)}_{i,l}$'s.

\subsection{Computation of Normalized Graph Edge Weights}

We next examine how we compute \textit{normalized} edge weights $\tilde{w}_{i,j}$'s to define GLR for the gradient graph $\bar{\cG}$.
Using CNN, we compute a low-dimensional \textit{feature vector} $\f_i \in \mathbb{R}^D$ for each node $i$ from embedding $\x_i \in \mathbb{R}^B$ as $\f_i = F(\x_i)$, where $D \ll B$. 
To compute \textit{unnormalized} edge weight $w_{i,j}$ between nodes $i$ and $j$, we first compute the squared Euclidean distance $d(i, j) = \|\f_i - \f_j\|^2_2$ in feature space, and $w_{i,j} = \exp \left( -d(i,j) \right)$. 
Then, \textit{normalized} edge weight $\tilde{w}_{i,j}$ is computed as
\begin{align}
\tilde{w}_{i,j} = \frac{w_{i,j}}{\sum_l w_{i,l}} = \frac{\exp \left(-d(i,j) \right)} {\sum_{l} \exp \left( -d(i,l) \right)} .
\label{eq:edgeWeight}
\end{align}
In matrix form, the \textit{random-walk graph Laplacian} is thus $\tilde{\L} = \bar{\D}^{-1} \bar{\L}$.
Since $\tilde{\L}$ is not symmetric, instead of conventional GLR, we define a variant of GLR based on the random-walk Laplacian \cite{liu17}: 
\begin{align}
\balpha^\top \tilde{\L}^\top \tilde{\L} \balpha = 
\x^\top \underbrace{\F^\top \tilde{\L}^\top \tilde{\L} \F}_{\tilde{\cL}} \x = \x^\top \tilde{\cL} \x .
\label{eq:GGLR2}
\end{align}
Unlike \eqref{eq:GGLR}, GGLR in \eqref{eq:GGLR2} is defined using a normalized Laplacian, and yet computes to zero for constant gradient---\ie, $\1^\top \tilde{\L}^\top \tilde{\L} \1 = 0$, as expected for a linear signal.


\vspace{0.05in}
\noindent 
\textbf{Comparison to Self-Attention Operator}: 
We see how the computation of edge weights \eqref{eq:edgeWeight} is akin to the computation of attention weights in \eqref{eq:SA_coeff2}.  
Specifically, \textit{interpreting the negative squared Euclidean distance $-d(i,j)$ as affinity $e(i,j)$, normalized edge weights $\tilde{w}_{i,j}$'s in \eqref{eq:edgeWeight} are essentially the same as attention weights $a_{i,j}$'s in \eqref{eq:SA_coeff2}}. 
However, $-d(i,j)$'s are computed using shallow CNNs that are more compact than dense query and key matrices, $\Q$ and $\K$. 
Further, given a graph $\cG$ specified by $\tilde{\cL}$, the restored signal $\x^*$ is a low-pass filtered $(\A^\top \A + \mu \tilde{\cL})^{-1}$ output of $\A^\top \y$ in linear system \eqref{eq:linSys} (computed via our proposed family of unrolled PnP-GGLR algorithms), while the output embedding $\y_i$ in a conventional transformer requires dense value matrix $\V$ in \eqref{eq:SA_operator}. 
\blue{Thus, each graph learning module corresponds to the learning of attention weights, while the GGLR-based optimization corresponds to parameter-efficient attention-based filtering.}



%% file: image/model.tex
\begin{figure*}[t]
\begin{center}
\centerline{\includegraphics[width=0.90\textwidth]{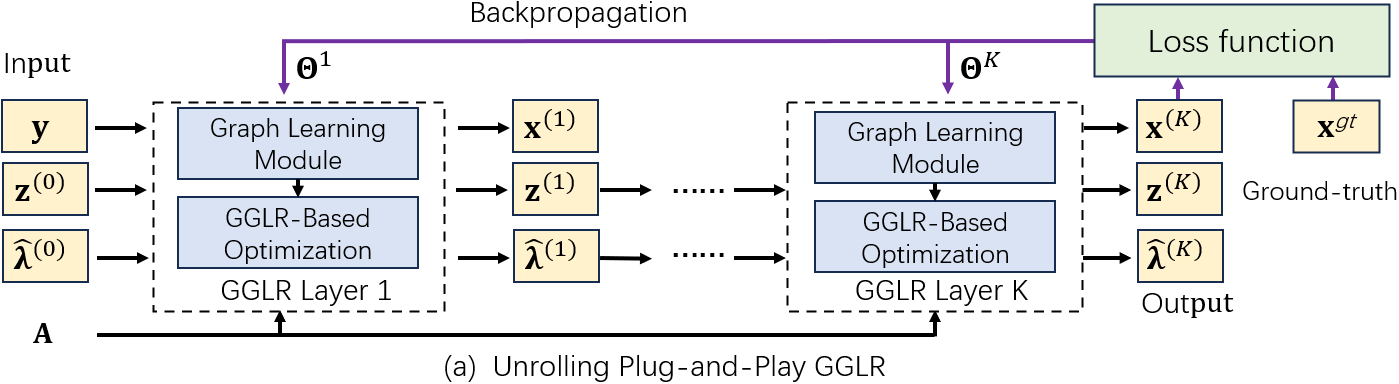}}
\centerline{\includegraphics[width=0.90\textwidth]{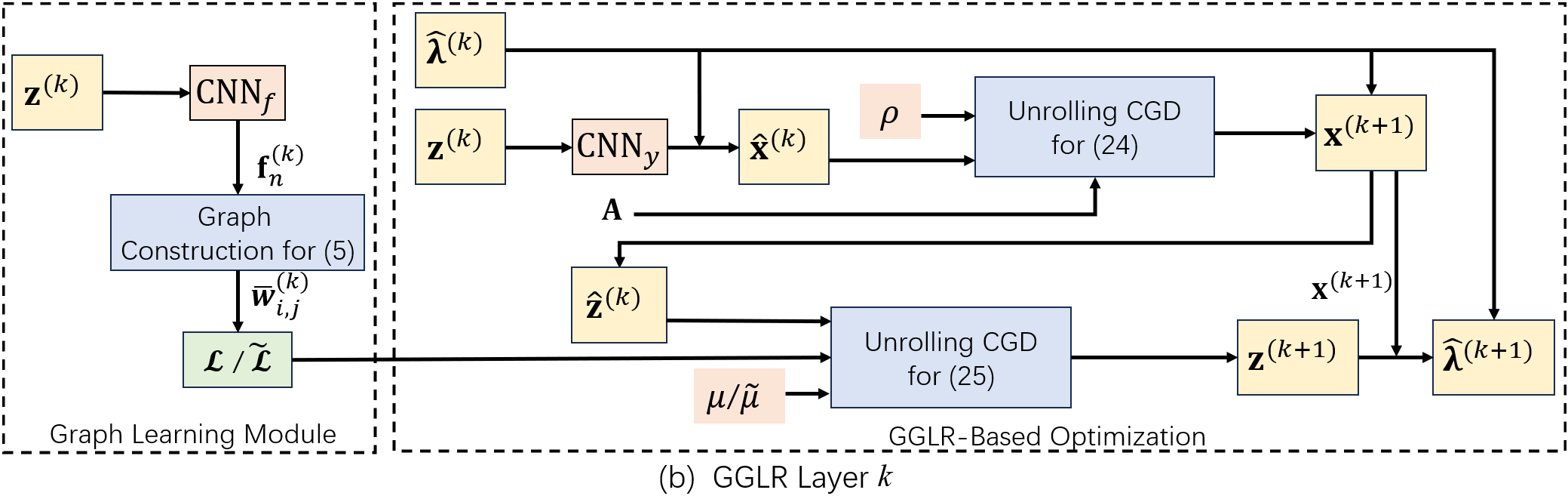}}
\centerline{\includegraphics[width=0.90\textwidth]{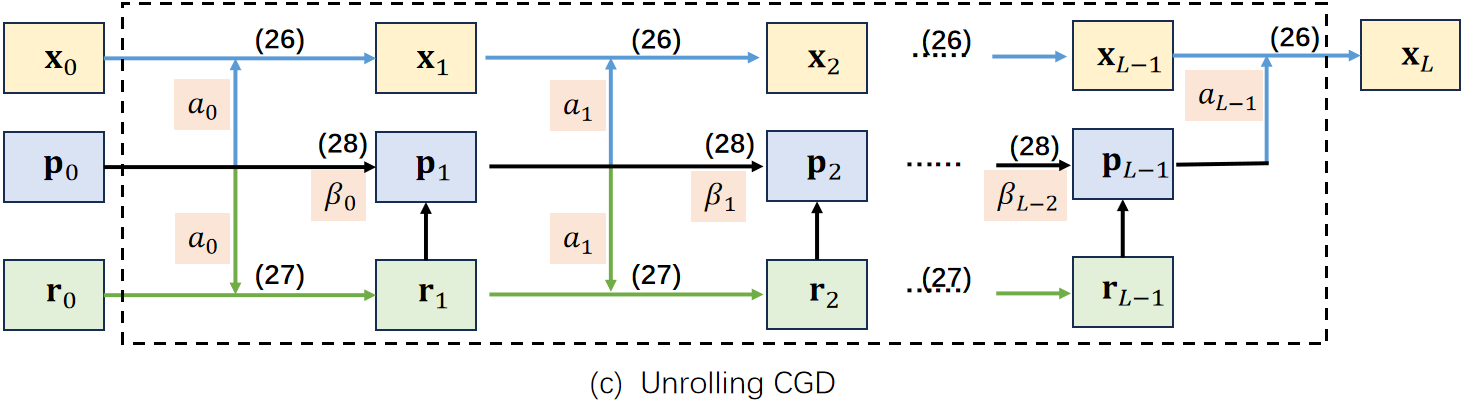}}
\end{center}
\vspace{-0.4in}
\caption{Overview of the proposed architecture. (a) UPnPGGLR composed of multiple GGLR Layers, (b) Block diagram of GGLR Layer, (c) Unrolling CGD.}
\label{fig:model}
\end{figure*}

%% file: results.tex
\input{image/awdn_denoise}

\subsection{Experimental Setup}
\label{sec:xperimental Setup}
We set the number of GGLR layers to $K=10$ and the number of CGD layers to $L=10$. 
The $\mathrm{CNN}_\mathrm{f}$ consists of $6$ convolution layers. 
The first layer has $3$ input channels and $32$ output channels. 
The last layer has $32$ input channels and $3$ output channels.
We use a $\operatorname{ReLU}(\cdot)$ activation function after every convolutional layer.
The $\mathrm{CNN}_\mathrm{y}$ consists of 4 convolutional layers using a residual learning structure.
For both training and testing, the patch size is set to $36\times36$, and the patches are extracted from images in an overlapped sliding window, where the stride is set to $32$. 
In each experiment, we train our proposed UPnPGGLR model for $200$ epochs using stochastic gradient descent (SGD). 
The batch size is set to $16$, and the learning rate is set to $1e^{-4}$. 
The proposed model is implemented in PyTorch and trained on an Nvidia 3090 GPU. 
We evaluate the performance using two common image metrics, peak signal-to-noise ratio (PSNR), and structural similarity index measure (SSIM) \cite{wang2004image}, on the test images. 

\blue{Parameters $\sigma_{a}$ and $\sigma_{f}$ in \eqref{eq:gradEdgeWeight} were initialized to $0.01$, parameters $\mu$ and $\tilde{\mu}$ were initialized to $0.3$, and parameter $\rho$ was initialized to $1$.
Constraints were imposed on $\mu$, $\tilde{\mu}$ and $\rho$ to ensure their non-negativity.
These constraints not only align with the mathematical contexts of different terms in the optimization objective, but also contribute to the stability
and robustness of the training process.
As the training progressed through iterations, these parameters eventually converged to fixed values. 
A detailed discussion is presented in Section\;\ref{sec: learnable Parameter Variation}.}

\blue{
We note that though in principle the same pre-trained GGLR-based denoiser can be reused for different settings, towards optimal performance, we retrained our model parameters for different definitions of $\A$, given that the statistics for different applications differ in general.
}

\blue{
To validate the importance of the graph learning module (periodic learning of attention weights), we introduced a variant named UPnPGGLR-S, where a single fixed (but optimized) graph is used for GGLR optimization across all layers. 
We will demonstrate that UPnPGGLR outperforms UPnPGGLR-S generally, thereby demonstrating the importance of the graph learning module in enhancing restoration performance.
}

\input{image/cross_domain.tex}

\input{table/awdn_denoise.tex}

\input{table/generalization}

\input{table/ablation}

\subsection{Image Denoising}
\label{sec:denoising}


For additive white Gaussian noises (AWGN) denoising, we set the degradation matrix $\A$ in \eqref{eq:formation} to identity. 
We used BSDS500 \cite{roth2009fields} for evaluation as done in \cite{DnCNN}. 
It consists of 432 training images and 68 test images. 
They were corrupted by AWGN of standard deviation $\sigma$ ranging from 5 to 55.

We first compare our proposed UPnPGGLR model with CBM3D \cite{BM3D}, TWSC \cite{xu18}, NSS \cite{hou20}, CDnCNN \cite{DnCNN}, IRCNN \cite{zhang2017learning}, FFDNet \cite{FFDNet}, DeepGLR \cite{deepglr}, DeepGTV \cite{deepgtv}, DRUNet \cite{zhang2021plug}, Restormer \cite{Zamir2021RestormerET}.  
Note that CBM3D, TWSC and NSS are pure model-based
methods, and thus do not require data-driven parameter tuning.
DeepGLR and DeepGTV are hybrid methods that combine CNNs with graph smoothness priors. Restormer is a method based on transformers, whereas DRUNet uses a pure CNN model as a denoiser. Both Restormer and DRUNet require a substantial number of network parameters and are prone to overfitting.

To evaluate the proposed method, we conducted comparative experiments at noise levels $\sigma = \{10, 25, 50\}$. The PSNR and SSIM results are shown in Table \ref{tab2}. 
While model-based methods TWSC \cite{xu18} and NSS \cite{hou20} are more interpretable, their denoising performances are glaringly inferior compared to SOTA DNNs, due to their overly restrictive assumed models.
We observe that Restormer achieves the best PSNR results on AWGN, but outperforming our proposed UPnPGGLR only slightly. 
However, Restormer incurs a significant larger number of parameters---UPnPGGLR employs less than $1\%$ of the parameters in Restormer. 
Compared to similar hybrid models such as DeepGLR and DeepGTV, UPnPGGLR achieves noticeably better performance. 
\blue{Further, UPnPGGLR achieves better performance than UPnPGGLR-S, demonstrating the importance of the graph learning module in denoising.}

A visual comparison of the models is shown in Fig. \ref{fig:awdn_denoise}. 
We observe that DeepGLR and DeepGTV fail to fully remove noises in the restored images, while UPnPGGLR obtains more visually satisfactory results---the restored images appear less blocky and more natural. 
Moreover, our network performs comparably to generic DNNs like DRUNet while using fewer than $1\%$ of the parameters.

\subsubsection{Robustness to Covariate Shift} 

The robustness of our approach is evaluated in terms of its cross-domain generalization ability. Specifically, the evaluation is conducted on the real noise dataset RENOIR \cite{renoir} and Nam-CC15 \cite{nam2016holistic}, using the model trained for AWGN of standard deviation $\sigma\in[5, 55]$.
The RENOIR dataset consists of 10 high-resolution real low-light noisy images and their corresponding noise-free versions. The Nam-CC15 dataset includes 15 noisy images of size $512\times512$ captured by Canon 5D Mark III, Nikon D600, and Nikon D800 cameras.

We compare our proposed UPnPGGLR model only with top-performing denoisers in Table\;\ref{tab2}:  CDnCNN \cite{DnCNN}, IRCNN \cite{zhang2017learning}, FFDNet \cite{FFDNet}, DeepGLR \cite{deepglr}, DeepGTV \cite{deepgtv}, DRUNet \cite{zhang2021plug} and Restormer \cite{Zamir2021RestormerET}. 
PSNR and SSIM results are shown in Table \ref{tab_generalization}. 
We observe that UPnPGGLR surpasses Restormer by 0.83dB PSNR on the RENOIR dataset and by 0.94dB PSNR on the Nam-CC dataset. 
However, Restormer provides better results in Gaussian noise removal, as we saw in Table\;\ref{tab2}. 
This indicates that Restormer is overfitted to Gaussian noise and fails to generalize to real noise, whereas UPnPGGLR provides satisfactory denoising results in both cases. 
Compared to similar hybrid models such as DeepGLR and DeepGTV, UPnPGGLR achieves the best PSNR results on both real image noise datasets. 
Fig.\;\ref{fig:cross} shows representative visual denoising results of the RENOIR and Nam-CC15 datasets. 
We observe that the competing methods fail to remove the real noise in the restored images, whereas the proposed method provides better visual quality.

Fig.\;\ref{fig:curve}(a) shows the performance of different methods under covariate shift. 
The models were trained with noise $\sigma=15$ on the BSDS500 dataset and tested at various noise levels. 
As the noise level increases, the performance of all models decreases. 
However, our method shows a noticeable improvement in robustness to covariate shift over competitors.


\input{image/interpolation/interpolation_more}

\input{image/curve/curve}
\subsubsection{Effects of Auxiliary Variables}
We compare different variants of UPnPGGLR algorithms with different number of auxiliary variables for different restoration scenarios: denoising, interpolation, and non-blind deblurring (the datasets will be introduced in the sequel). 
\blue{
In Table\;\ref{table_ablation}, GGLR denotes the model of solving the linear system \eqref{eq:linSys} directly by unrolling CGD, while UPnPGGLR-O, UPnPGGLR-T, and UPnPGGLR-F denote the ADMM algorithm with one auxiliary variable ($\z$), two auxiliary variables ($\z_r$, $\z_c$), and four auxiliary variables ($\z_r$, $\z_c$, $\tilde{\z}_r$, $\tilde{\z}_c$), respectively.
The first and second rows in Table\;\ref{table_ablation} show the performance of DeepGLR and DeepGTV in image restoration.
We see that the performance of GGLR is superior to that of DeepGLR and DeepGTV. 
From the fourth to sixth rows of the table, we see a progressive improvement in performance by gradually introducing more auxiliary variables.
This improvement is attributed to the network's ability to learn more parameters along individual dimensions.
Note that introducing additional auxiliary variables only marginally increases the parameter count. 
Further, ADMM induces a structured and interpretable optimization process
by decoupling the problem into simpler sub-problems focusing on minimization of individual terms. 
This structured approach not only enhances the clarity of the optimization process but
also contributes to the overall effectiveness of the proposed method.
}

\input{table/table_flops}

\subsubsection{Ablation Study}
We investigate the impact of the number of GGLR layers and the iteration numbers of CGD.
Experiments are conducted on the RENOIR dataset \cite{renoir} following recent works \cite{deepglr, deepgtv}. The images are split into two sets for training and testing, with each set containing five images.
Fig.\;\ref{fig:curve}(b) shows the average PSNR results of the proposed UPnPGGLR with different GGLR layer numbers $K$, given a fixed CGD iteration number $L=10$.
It is evident that PSNR of UPnPGGLR increases rapidly for $K\leq10$, but the improvement becomes marginal for $K > 10$.
Fig.\;\ref{fig:curve}(c) shows the influence of the iteration numbers of GCD given $K=10$.
Similarly, we can see that the PSNR performance improves as the number of iterations increases. 
Ultimately, we set $K=10$ and $L=10$.
\blue{Additionally, we find that while increasing the number of auxiliary variables improves model performance, it has minimal impact on the selection of layers and iterations.}

\subsubsection{Computational Efficiency}

\blue{We conducted experiments to compare the computational efficiency of different methods in terms of parameter count, runtime (inference time), and FLOPs. 
The results are summarized in Table\;\ref{table_flop}. 
All experiments were performed on the CBSD68 denoising dataset with images of size $480 \times 320$. 
Our method achieves a good tradeoff between FLOPS (10.43 G) and parameter count (0.23 M). 
In contrast, Restormer and DRUNet require substantial parameter counts and computational resources (FLOPS).
}

\blue{
We note that runtimes for graph-based deep unrolling schemes (\eg, DeepGLR, DeepGTV, UPnPGGLR) are noticeably larger. 
This is due to the current lack of efficient
implementations of sparse matrix-vector multiplication. 
Addressing this software engineering challenge to improve runtime is left for future research.
}


\subsection{Image Interpolation}

\input{table/interpolation_table}

For image interpolation, the degradation matrix $\A$ in \eqref{eq:formation} is a sampling matrix.
We conducted training on the 432 images of the BSDS500 dataset \cite{roth2009fields} and tested on the Set5 dataset \cite{timofte2015a+}. 
We compared our method against SOTA image interpolation techniques, including EPLL \cite{zoran2011learning}, IRCNN \cite{zhang2017learning}, DeepGLR \cite{deepglr}, DeepGTV \cite{deepgtv} \blue{and the latest plug-and-play method, SRONE \cite{SNORE}.}
The average PSNR and SSIM of the reconstructed images are shown in Table\;\ref{table_interpolation}. 
Notably, UPnPGGLR demonstrates superior performance compared to the competing methods, particularly in scenarios with a large fraction of missing pixels.
\blue{For instance, in the case of 80\% pixel loss, UPnPGGLR outperforms SNORE by achieving a PSNR of 28.98 dB.}
A representative visual comparison is shown in Fig.\;\ref{fig:visual_interpolation}. 
We observe that IRCNN tends to over-smooth edges, while our method can reconstruct images with sharp edges and natural textures.
\textit{Note that UPnPGGLR employs fewer than $5\%$ of the parameters in IRCNN}.


\input{image/deblurry/deblur/deblur_more}

\subsection{Non-Blind Image Deblurring}

For non-blind image deblurring, degredation matrix $\A$ in \eqref{eq:formation} is a known low-pass blur filter. 
The blurry images were synthesized by first applying a blur kernel and then adding additive Gaussian noise with noise level $\sigma$.
For training, we filtered $432$ images from BSDS500 \cite{roth2009fields} with the two real blur kernels from \cite{levin2009understanding} and then added 1\% white Gaussian noise to the images. 
We created a test dataset Set6 which selected six images from \cite{zhang2017learning}.

To evaluate the effectiveness of UPnPGGLR, we compared it against four representative methods: the model-based EPLL \cite{zoran2011learning} and RGTV \cite{bai19}, the learning-based non-blind FDN \cite{kruse2017learning} and IRCNN \cite{zhang2017learning}, \blue{and recent plug-and-play approaches Eq.DRUNet \cite{EqDRUNtet} and SNORE \cite{SNORE}.} 
The PSNR and SSIM of different methods are shown in Table\;\ref{table_deblur}. 
\blue{Our method achieves superior results compared to
both Eq.DRUNet and SNORE across two different blur kernels.}
Fig.\;\ref{fig:visual_deblur} visually compares deblurred images by different methods. 
We observe that RGTV and FDN tend to smooth out fine details and generate color artifacts.
Although IRCNN avoids color artifacts, it fails to recover fine details.
In contrast, the proposed method can recover image edges and textures more naturally.
\input{table/deblurry_table.tex}
\input{image/parameter}

\subsection{Learnable Parameter Variation}
\label{sec: learnable Parameter Variation}

\blue{
The learnable parameters in our work are mainly $\{\sigma_{a}^{(k)}, \sigma_{f}^{(k)}, \rho^{(k)},\mu^{(k)}, \tilde{\mu}^{(k)},  
\mathbf{\alpha}_\mathbf{x}^{(k)},\mathbf{\beta}_\mathbf{x}^{(k)}, \mathbf{\alpha}_\mathbf{z}^{(k)},\mathbf{\beta}_\mathbf{z}^{(k)}\}$. 
We study the evolution of these learned parameters for further insights.
We described the initialization of key learnable parameters in Section\;\ref{sec:xperimental Setup}.  
Given the initial values, these parameters gradually converge to specific values during the training process. 
Fig.\;\ref{fig:pamameter} shows the evolution of parameters $\mu$, $\tilde{\mu}$, and $\rho$ across iterations for the first GGLR layer. 
For instance, the initial values of $\mu$ and $\tilde{\mu}$ were set to $0.3$, and both eventually converged to near 1. 
These parameters control the weights of the horizontal and vertical GGLR terms, respectively; their converged values are similar, reflecting their balanced contributions to the model. }


%% file: image/awdn_denoise.tex

\begin{figure*}[t]
    \centering
    \footnotesize
	\begin{tabular}{c@{}c@{}c@{}c@{}c@{}c}
        \includegraphics[width=2.90cm] 
        {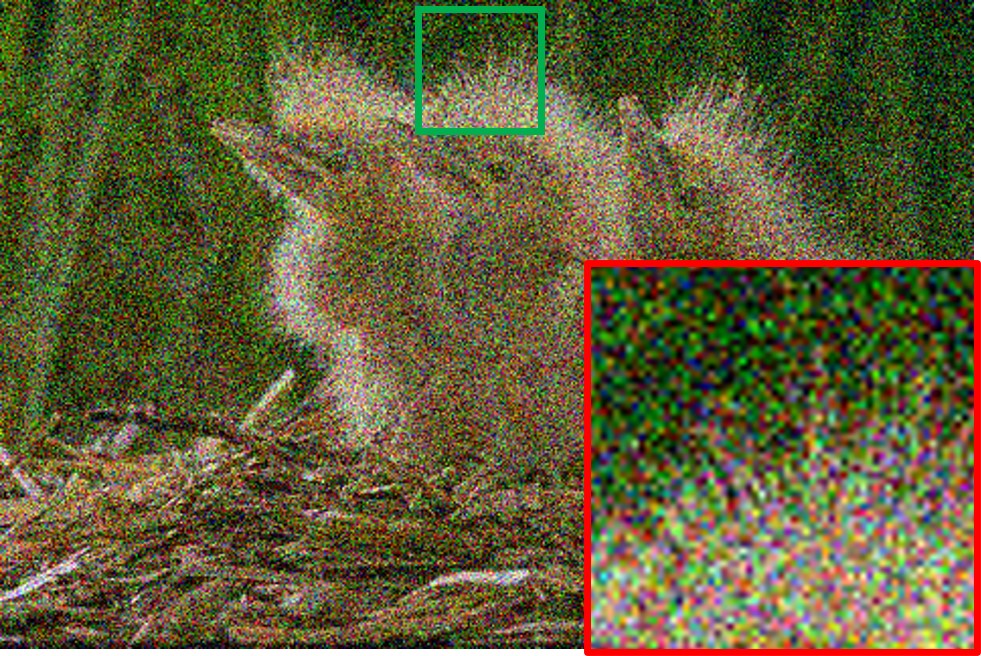} \hspace{-0.04mm} & 
		  \includegraphics[width=2.90cm] 
        {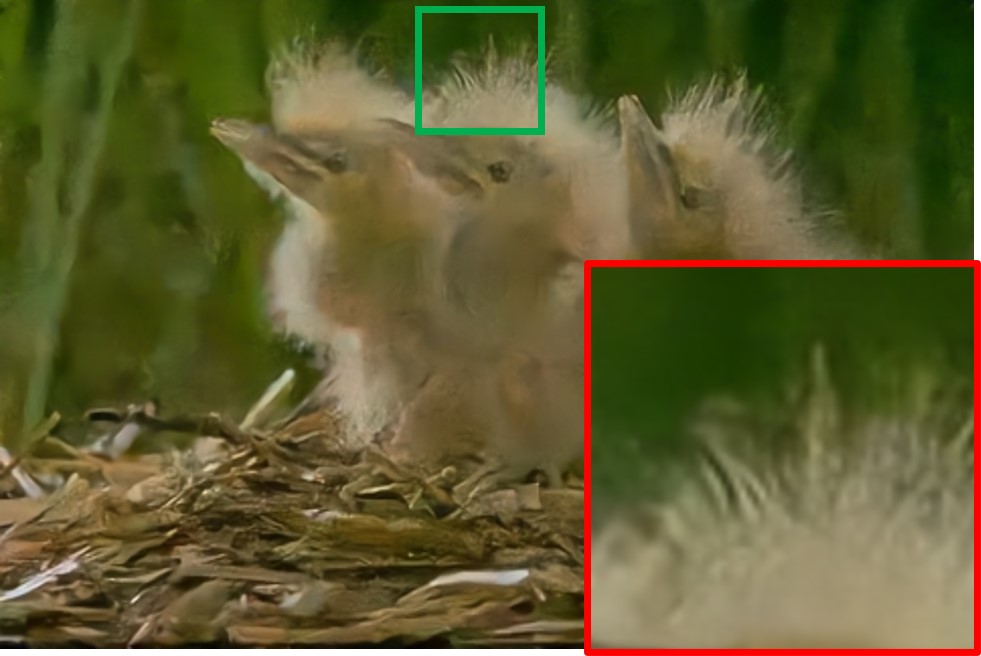} \hspace{-0.04mm} & 
		  \includegraphics[width=2.90cm] 
        {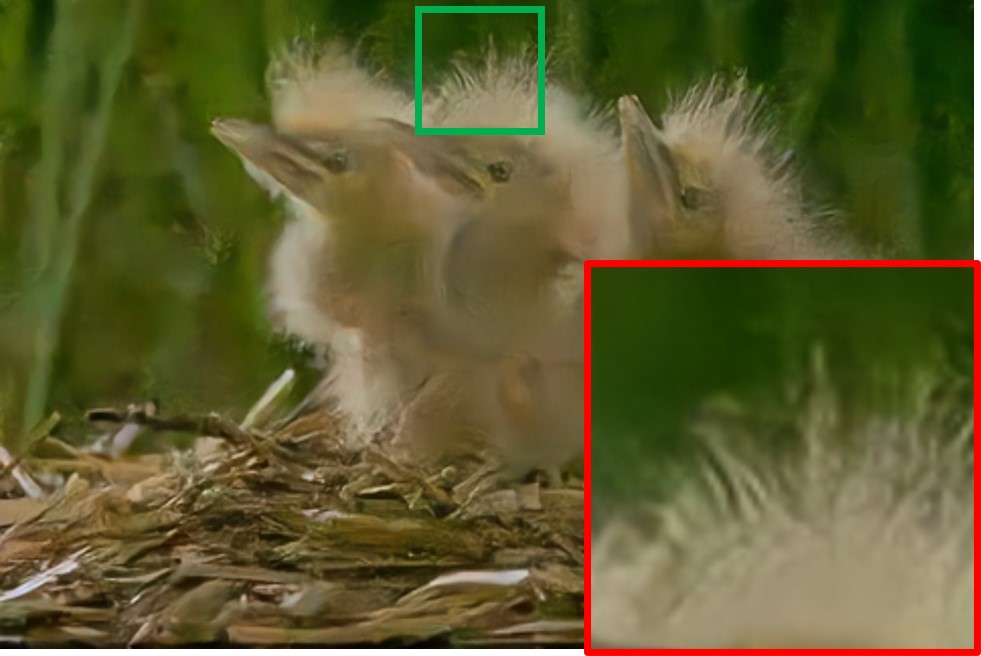} \hspace{-0.04mm} & 
          \includegraphics[width=2.90cm]
       {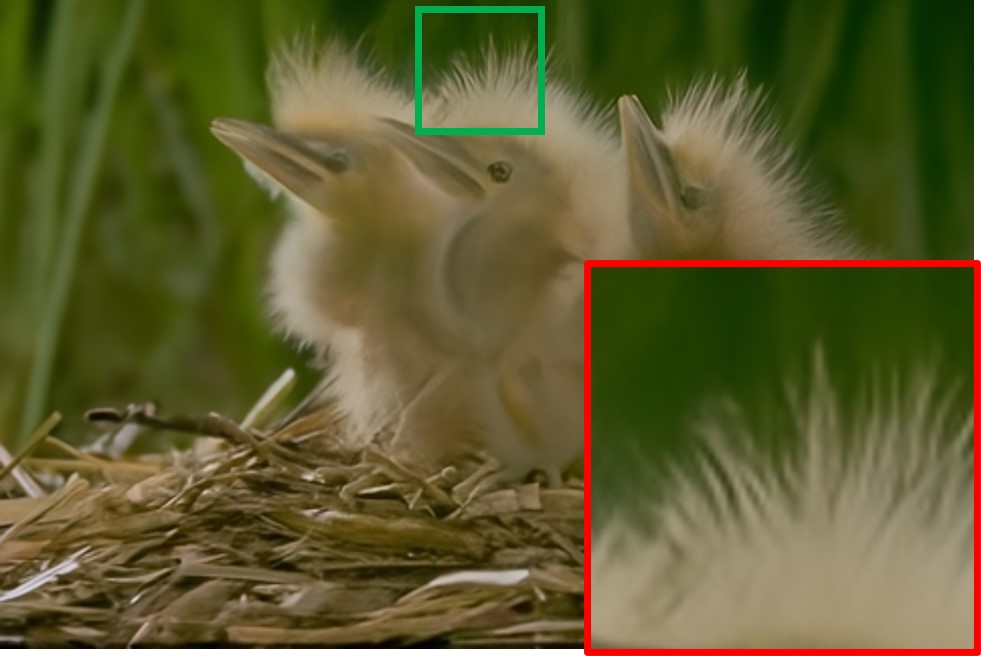} \hspace{-0.04mm} & 
          \includegraphics[width=2.90cm] 
        {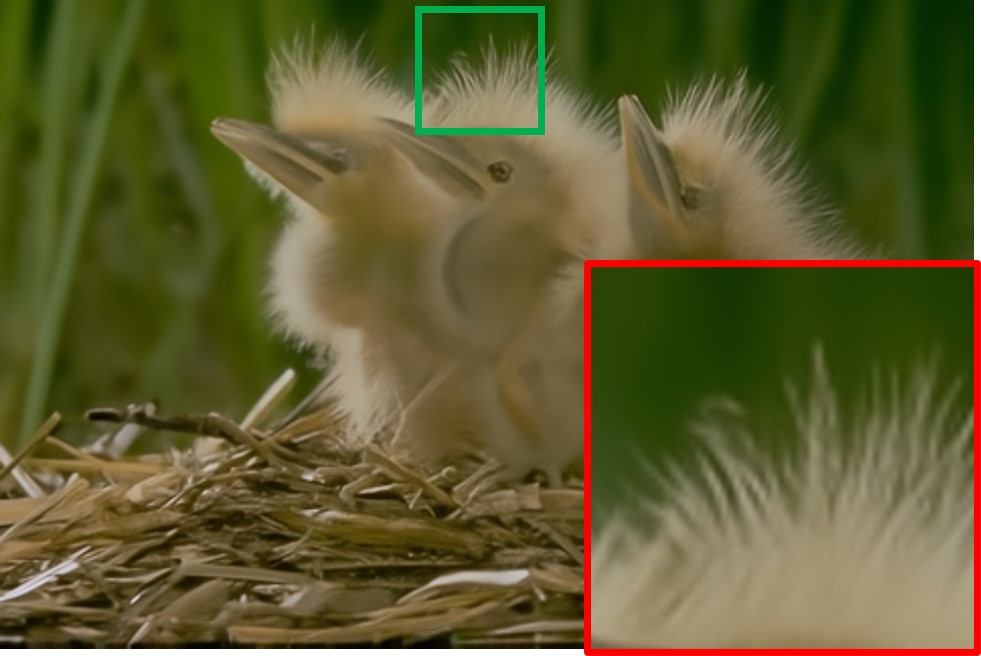} \hspace{-0.04mm} & 
          \includegraphics[width=2.90cm] 
        {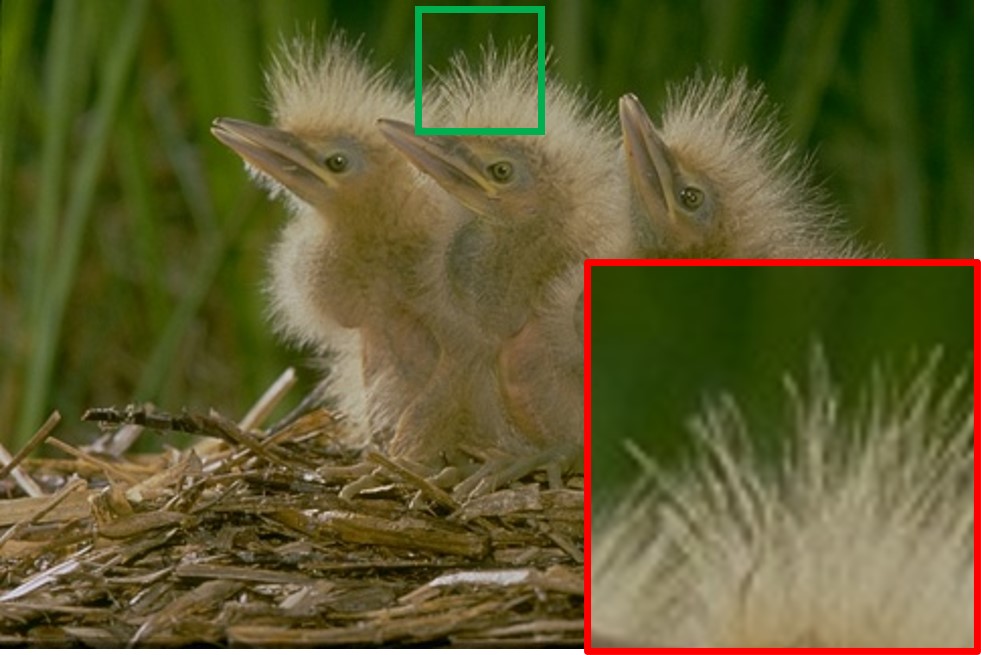} \hspace{-0.04mm} \\
        (a) Noisy / 14.99dB & (b) DeepGLR / 28.72dB & (c) DeepGTV / 28.75dB & (d) DRUNet / 29.28dB & (e) UPnPGGLR / 29.24dB & (f) Ground-truth    \\

        \includegraphics[width=2.90cm] 
        {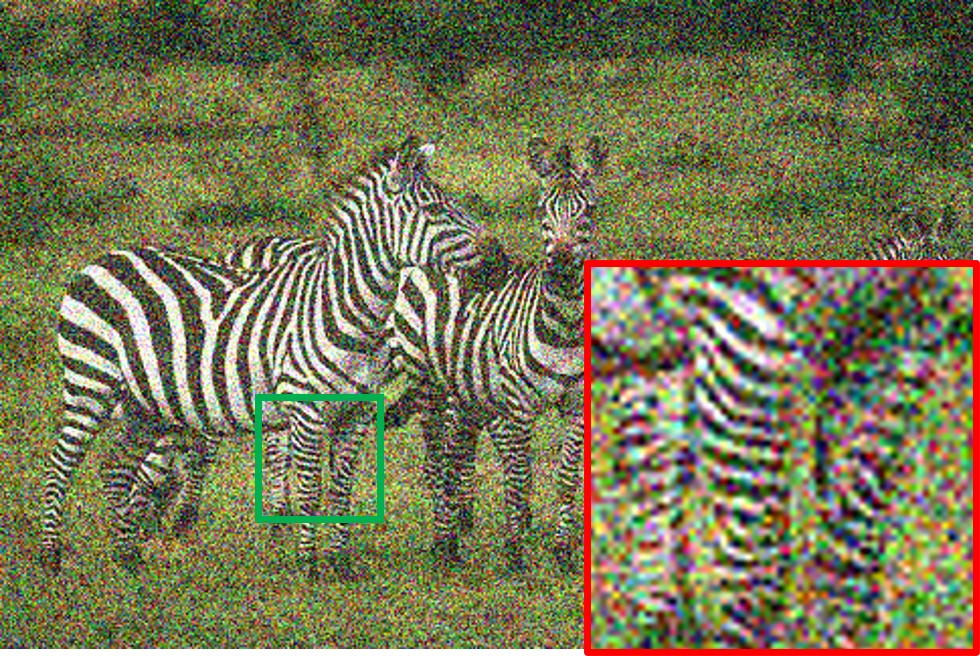} \hspace{-0.04mm} & 
		  \includegraphics[width=2.90cm] 
        {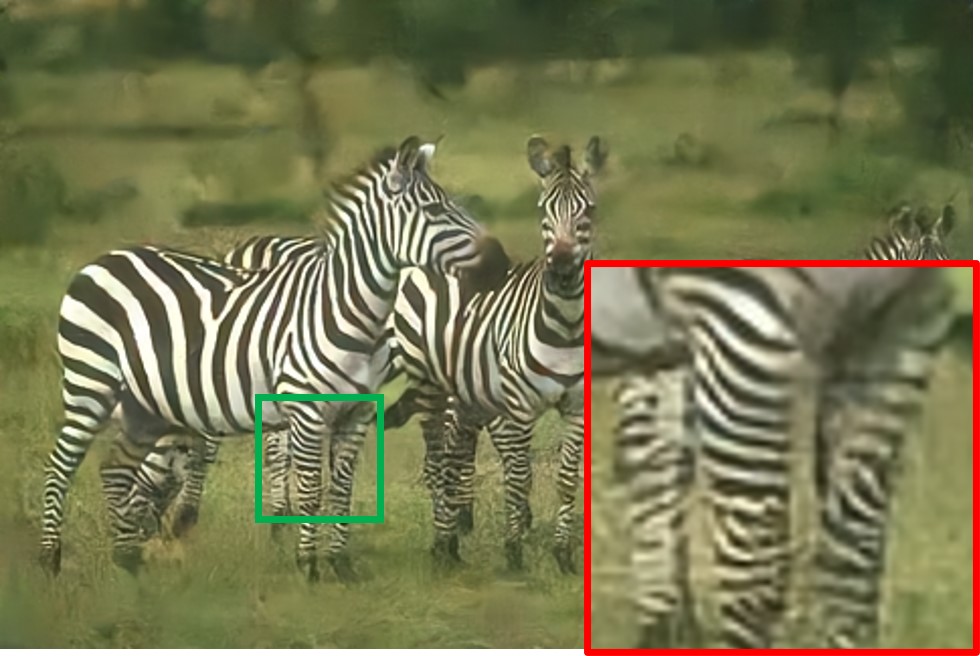} \hspace{-0.04mm} & 
		  \includegraphics[width=2.90cm] 
        {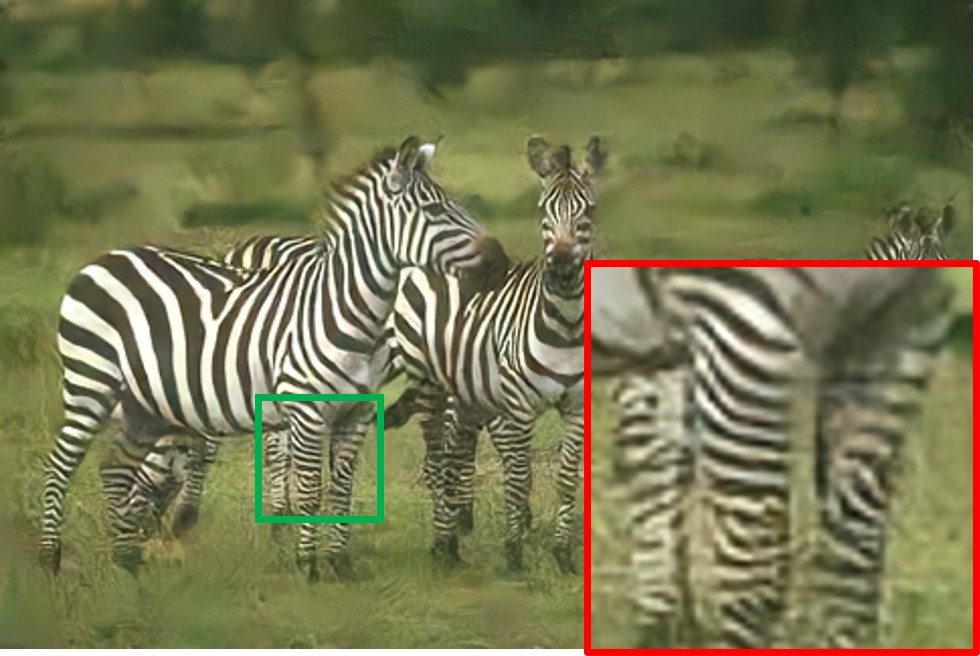} \hspace{-0.04mm} & 
          \includegraphics[width=2.90cm]
        {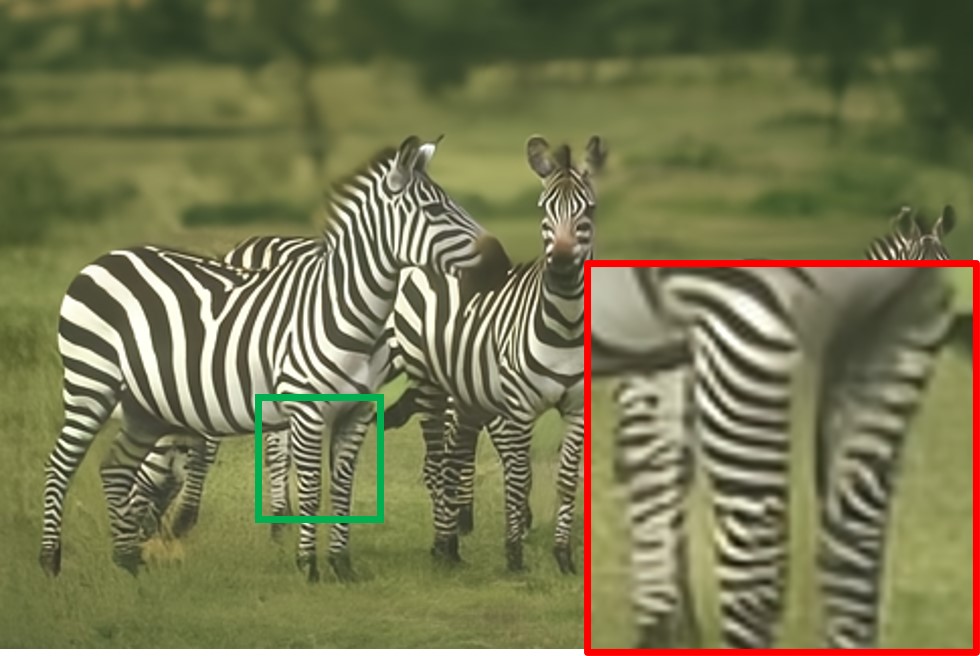} \hspace{-0.04mm} & 
          \includegraphics[width=2.90cm] 
        {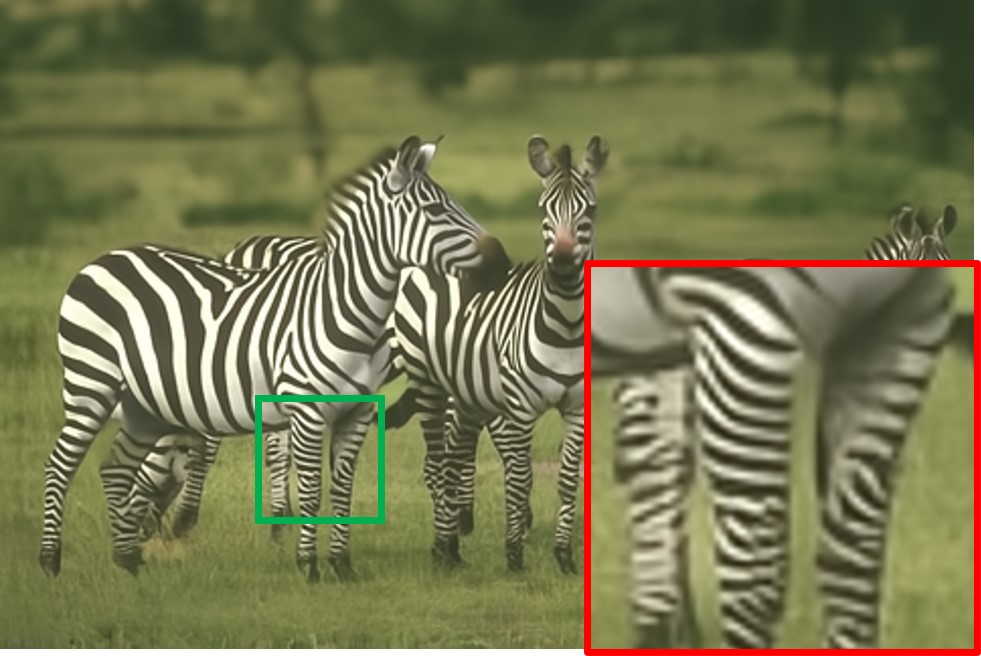} \hspace{-0.04mm} & 
          \includegraphics[width=2.90cm] 
        {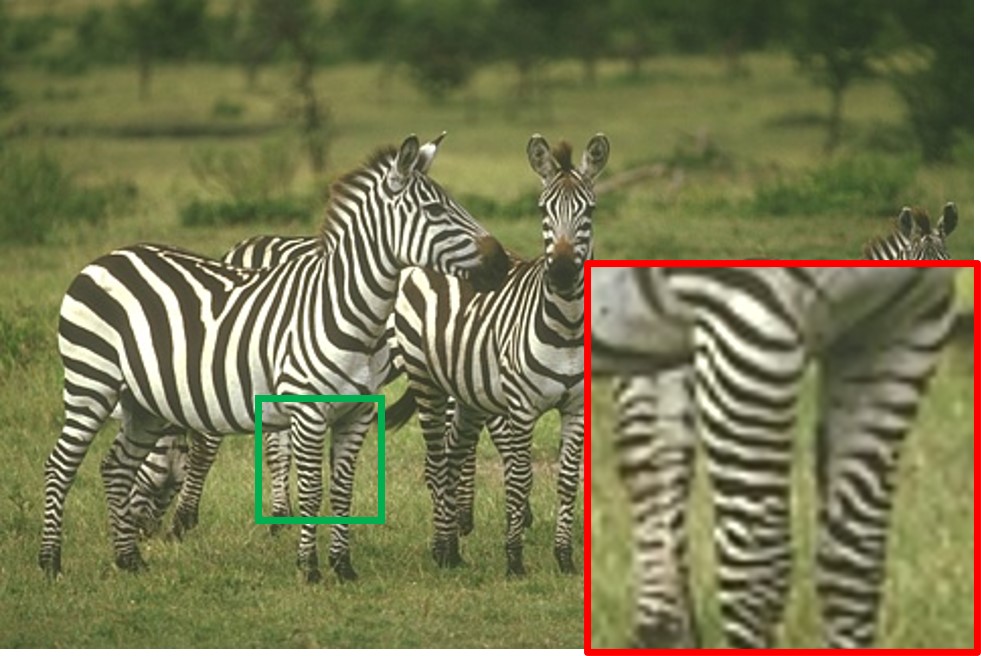} \hspace{-0.04mm} \\

         (a) Noisy / 14.99dB & (b) DeepGLR / 27.36dB & (c) DeepGTV / 27.54dB & (d) DRUNet / 28.07dB & (e) UPnPGGLR / 28.02dB & (f) Ground-truth

	\end{tabular}
	
	\caption{Color image denoising results of different methods on image “163085” and image "253027" from CBSD68 dataset with noise level 50.}

	\label{fig:awdn_denoise}
\end{figure*}

%% file: image/cross_domain.tex
\begin{figure*}[t]
  \centering
    \footnotesize
\begin{tabular}{c@{}c@{}c@{}c@{}c@{}c}
        \includegraphics[width=2.90cm] 
        {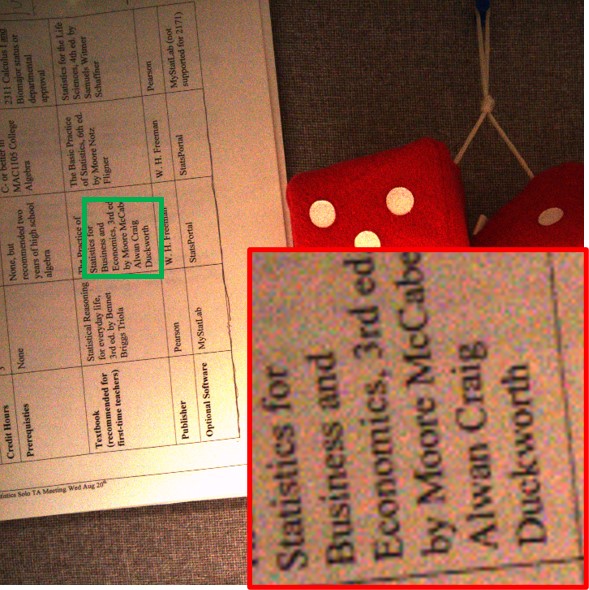} \hspace{-0.04mm} & 
		  \includegraphics[width=2.90cm] 
        {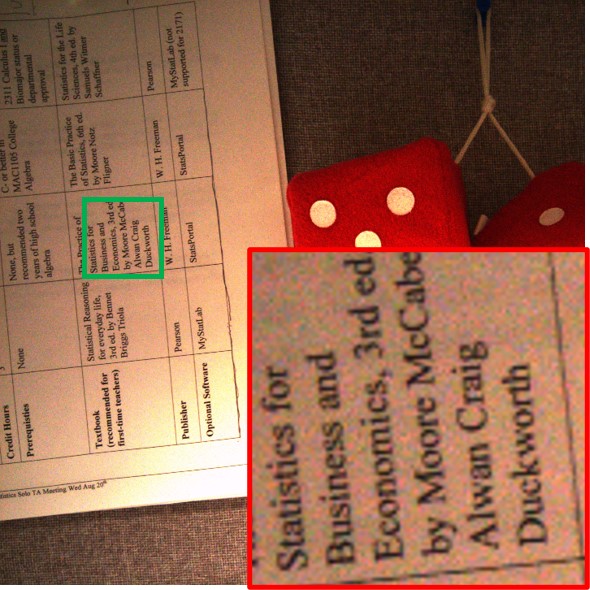} \hspace{-0.04mm} & 
		  \includegraphics[width=2.90cm] 
        {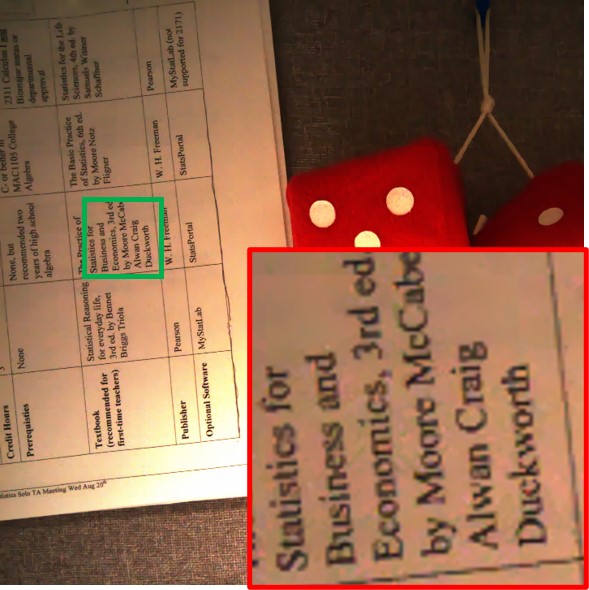} \hspace{-0.04mm} & 
          \includegraphics[width=2.90cm]
       {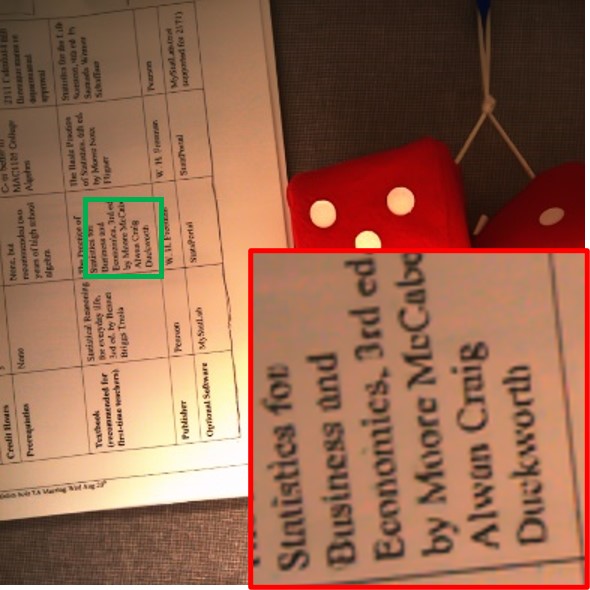} \hspace{-0.04mm} & 
          \includegraphics[width=2.90cm] 
        {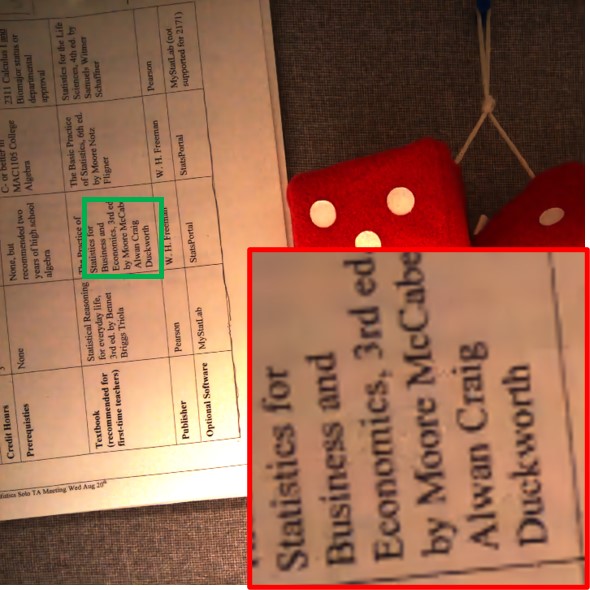} \hspace{-0.04mm} & 
          \includegraphics[width=2.90cm] 
        {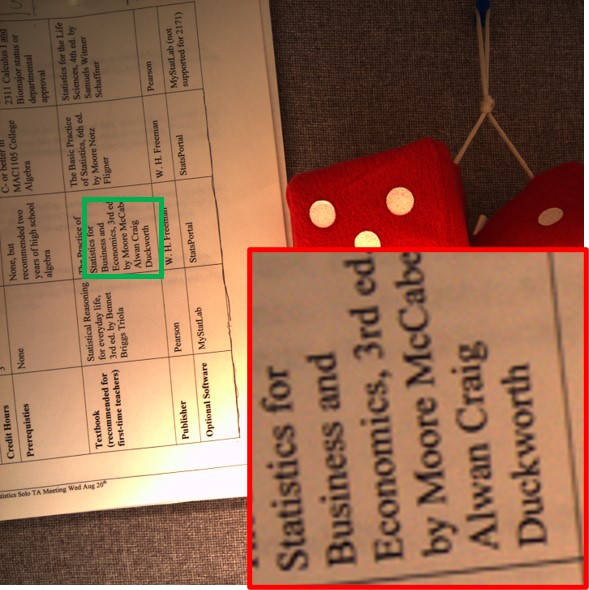} \hspace{-0.04mm} \\
        (a) Noisy / 23.39dB & (b) CDnCNN / 27.12dB & (c) DeepGLR / 29.78dB & (d) Restormer / 29.45dB & (e) UPnPGGLR / 30.01dB & (f) Ground-truth    \\

        \includegraphics[width=2.90cm] 
        {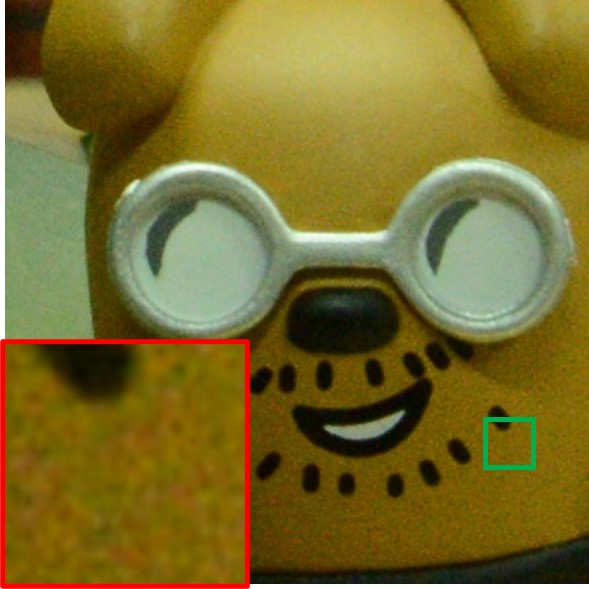} \hspace{-0.04mm} & 
		  \includegraphics[width=2.90cm] 
        {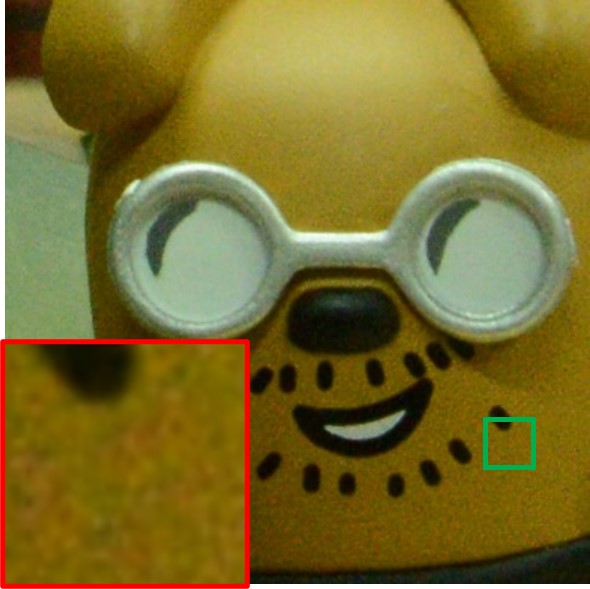} \hspace{-0.04mm} & 
		  \includegraphics[width=2.90cm] 
        {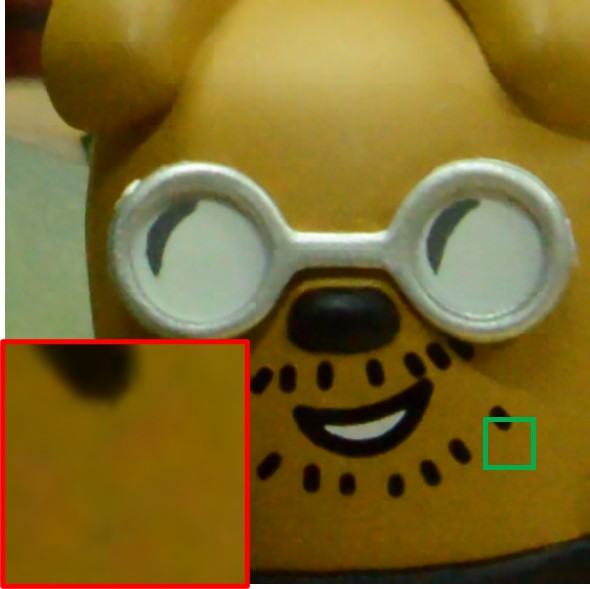} \hspace{-0.04mm} & 
          \includegraphics[width=2.90cm]
        {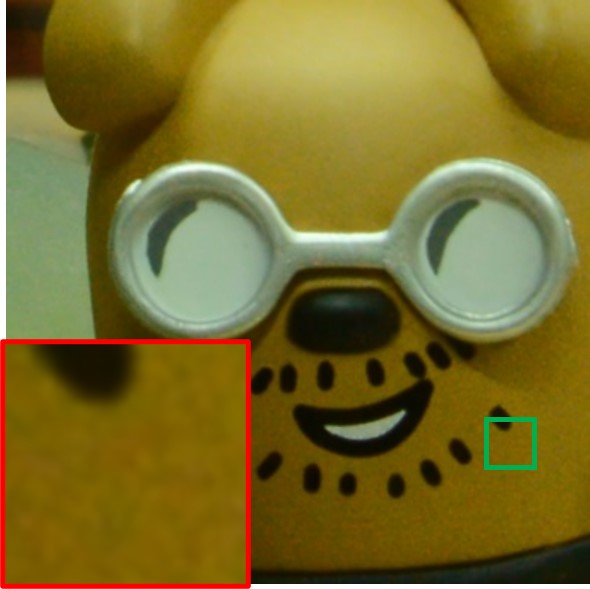} \hspace{-0.04mm} & 
          \includegraphics[width=2.90cm] 
        {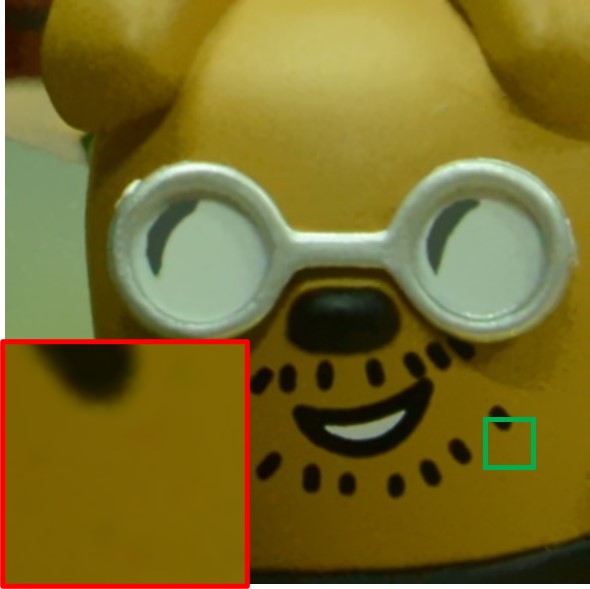} \hspace{-0.04mm} & 
          \includegraphics[width=2.90cm] 
        {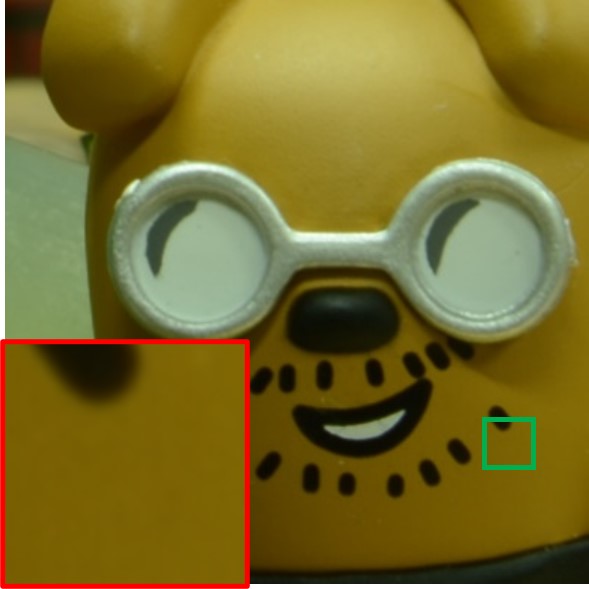} \hspace{-0.04mm} \\

         (a) Noisy / 33.51dB & (b) CDnCNN / 34.42dB & (c) DeepGLR / 38.39dB & (d) Restormer / 36.82dB & (e) UPnPGGLR / 39.01dB & (f) Ground-truth

	\end{tabular}
\vspace{-0.2cm}
\caption{Visual comparison of cross-domain generalization. Note that the proposed UPnPGGLR is trained on Gaussian white noises for image denoising and generalizes well to real image denoising.}
\label{fig:cross}
 \vspace{-0.1in}
\end{figure*}

%% file: table/awdn_denoise.tex
\begin{table}[t]
\caption{
Average PSNR(dB) results of different methods for noise levels $15$, $25$, and $50$ on CBSD68 \cite{roth2009fields}  dataset.}
\vspace{-0.3in}
\label{tab2}
\begin{center}    
\setlength{\tabcolsep}{3pt}
\resizebox{1.0\columnwidth}{!}{
\begin{tabular}{c|c|ccc}
\hline
\multirow{2}{*}{Method} & Paras(M) & \multicolumn{3}{c}{CBSD68 \cite{roth2009fields}}                                                        \\ \cline{2-5} 
                        & -        & \multicolumn{1}{c|}{$\sigma=15$}          & \multicolumn{1}{c|}{$\sigma=25$}          & $\sigma=50$         \\ \hline
CBM3D \cite{BM3D}                 & -     & \multicolumn{1}{c|}{33.49/0.922} & \multicolumn{1}{c|}{30.68/0.867} & 27.35/0.763 \\
TWSC \cite{xu18}                 & -     & \multicolumn{1}{c|}{33.41/0.918} & \multicolumn{1}{c|}{30.64/0.867} & 27.43/0.763 \\
NSS \cite{hou20}                 & -     & \multicolumn{1}{c|}{33.33/0.918} & \multicolumn{1}{c|}{30.76/0.868} & 27.61/0.769 \\
CDnCNN \cite{DnCNN}                 & 0.56     & \multicolumn{1}{c|}{33.90/0.929} & \multicolumn{1}{c|}{31.22/0.883} & 27.95/0.790 \\
IRCNN  \cite{zhang2017learning}                  & 4.75     & \multicolumn{1}{c|}{33.87/0.928} & \multicolumn{1}{c|}{31.18/0.882} & 27.86/0.789 \\
FFDNet \cite{FFDNet}                 & 0.67     & \multicolumn{1}{c|}{33.85/0.929} & \multicolumn{1}{c|}{31.22/0.883} & 27.98/0.792 \\
DeepGLR \cite{deepglr}                 & 0.93     & \multicolumn{1}{c|}{33.75/0.926} & \multicolumn{1}{c|}{31.13/0.881} & 27.86/0.791 \\
DeepGTV \cite{deepgtv}                & 0.10     & \multicolumn{1}{c|}{33.80/0.927} & \multicolumn{1}{c|}{31.08/0.879} & 27.90/0.791 \\
DRUNet \cite{zhang2021plug}                 & 32.64    & \multicolumn{1}{c|}{34.30/0.934} & \multicolumn{1}{c|}{31.69/0.893} & 28.51/0.810 \\
Restormer \cite{Zamir2021RestormerET}                & 26.11    & \multicolumn{1}{c|}{\textbf{34.39/0.935}} & \multicolumn{1}{c|}{\textbf{31.78/0.894}} & \textbf{28.59/0.813} \\
\blue{UPnPGGLR-S}                & \blue{\textbf{0.09}}    & \multicolumn{1}{c|}{\blue{33.91/0.930}} & \multicolumn{1}{c|}{\blue{31.19/0.882}} & \blue{27.95/0.792} \\ 
UPnPGGLR                & 0.23     & \multicolumn{1}{c|}{34.15/0.932} & \multicolumn{1}{c|}{31.42/0.889} & 28.18/0.802 \\ \hline
\end{tabular}
}
\end{center}
\vspace{-0.30in}
\end{table}

%% file: table/generalization.tex
\begin{table*}[t]
\caption{
Evaluation of cross-domain generalization for real image denoising on RENOIR \cite{renoir} and Nam-CC15 \cite{nam2016holistic} datasets. The best results are highlighted in boldface.}
\vspace{-0.2in}
\label{tab_generalization}
\begin{center}    
\large
\resizebox{0.95\linewidth}{!}{
\begin{tabular}{c|c|ccccccccc}
\hline
Dataset                   & Method & CDnCNN \cite{DnCNN} & IRCNN \cite{zhang2017learning}& FFDNet \cite{FFDNet} & DeepGLR \cite{deepglr} & DeepGTV \cite{deepgtv} & DRUNet \cite{zhang2021plug} & Restormer \cite{Zamir2021RestormerET}& \blue{UPnPGGLR-S} & UPnPGGLR \\ \hline
\multirow{2}{*}{RENOIR \cite{renoir}}   & PSNR   & 26.79  & 25.77 & 29.19  & 30.25   & 30.14   & 29.81  & 29.55   & \blue{30.26}  & \textbf{30.38}    \\
                          & SSIM   & 0.638  & 0.651 & 0.762  & 0.809   & 0.808   & 0.784  & 0.780     &  \blue{0.809} &\textbf{0.814}    \\ \hline
\multirow{2}{*}{Nam-CC15 \cite{nam2016holistic} } & PSNR   & 33.86  & 35.68 & 34.00  & 35.85   & 35.98   & 35.52  & 35.32  & \blue{36.06}  & \textbf{36.26}    \\
                          & SSIM   & 0.864  & 0.932 & 0.906  & 0.935   & 0.932   & 0.927  & 0.921  & \blue{0.934}   & \textbf{0.939}    \\ \hline
\end{tabular}
}
\end{center}
\vspace{-0.15in}
\end{table*}

%% file: table/ablation.tex
\begin{table*}[t]
\caption{\blue{The impact of the number of auxiliary variables on model performance.}}
\vspace{-0.2in}
\label{table_ablation}
\begin{center}
\setlength{\tabcolsep}{3pt}
\resizebox{0.85\linewidth}{!}{
\begin{tabular}{c|c|ccc|ccc|cc}
\hline
\multirow{2}{*}{\begin{tabular}[c]{@{}c@{}}Auxiliary\\ variables\end{tabular}}  & \multirow{2}{*}{\begin{tabular}[c]{@{}c@{}}\blue{Paras(M)}\end{tabular}} & \multicolumn{3}{c|}{AWDN denoising on CBSD68 \cite{roth2009fields}}   & \multicolumn{3}{c|}{Interpolation on set5 \cite{timofte2015a+}}      & \multicolumn{2}{c}{Deblurring on set6 \cite{zhang2017learning}} \\ \cline{3-10} 
& & $\sigma=15$ & $\sigma=25$ & $\sigma=50$           & 20\%        & 50\%        & 80\%        & Kernel 1       & Kernel 2      \\ \hline
 DeepGLR \cite{deepglr} & \blue{0.93}   & 33.75/0.926& 31.13/0.881& 27.86/0.791 & 38.92/0.981 & 33.75/0.948 & 27.48/0.843 & 33.28/0.924    & 33.53/0.928   \\
 DeepGTV \cite{deepgtv}& \bfseries{\blue{0.10}}   & 33.80/0.927& 31.08/0.879& 27.90/0.791 & 38.85/0.980 & 33.65/0.948 & 27.42/0.842 & 34.17/0.923    & 33.42/0.928   \\
GGLR & \blue{0.23}  & 33.92/0.929& 31.27/0.883& 28.01/0.792 & 39.61/0.982 & 34.53/0.951 & 28.62/0.869 & 34.38/0.937    & 34.04/0.934   \\
UPnPGGLR-O & \blue{0.23}  & 34.08/0.931& 31.35/0.888 &28.11/0.801 & 39.79/0.983 & 34.72/\bfseries{0.952} & 28.87/0.874 & 34.51/0.940    & 34.12/0.936   \\
UPnPGGLR-T & \blue{0.23} & 34.12/0.931  & 31.38/\bfseries{0.889}  & 28.15\bfseries{/0.802}           & 39.84/0.983           & 34.78/\bfseries{0.952}           & 28.94/0.876           & 34.54/\bfseries{0.941}              & 34.15/0.936             \\ 
UPnPGGLR-F & \blue{0.23} &   \bfseries{34.15/0.932}&   \bfseries{31.42/0.889}&   \bfseries{28.18/0.802}&  \bfseries{39.88/0.984} &  \bfseries{34.81/0.952} & \bfseries{28.98/0.877}&  \bfseries{34.57/0.941}&  \bfseries{34.18/0.937} \\
\hline
\end{tabular}
}
\vspace{-0.2in}
\end{center}
\end{table*}

%% file: image/interpolation/interpolation_more.tex
\begin{figure*}[t]
    \centering
    \footnotesize
   \fontsize{6}{10}\selectfont
	\begin{tabular}{c@{}c@{}c@{}c@{}c@{}c}

           \includegraphics[width=2.8cm] 
        {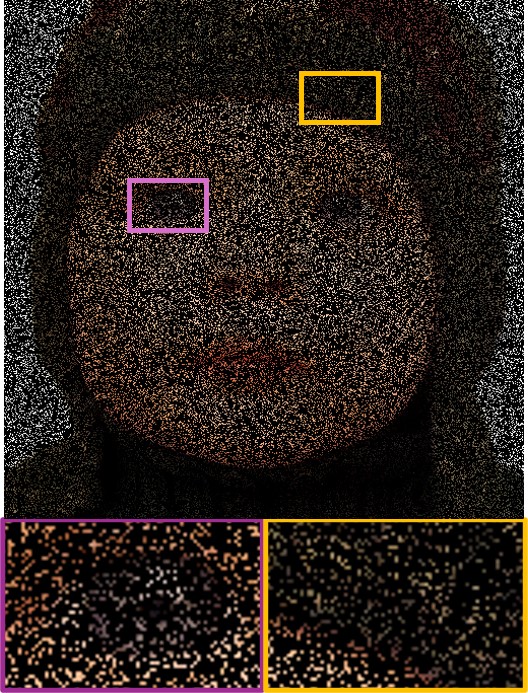} \hfill & 
         \includegraphics[width=2.8cm] 
        {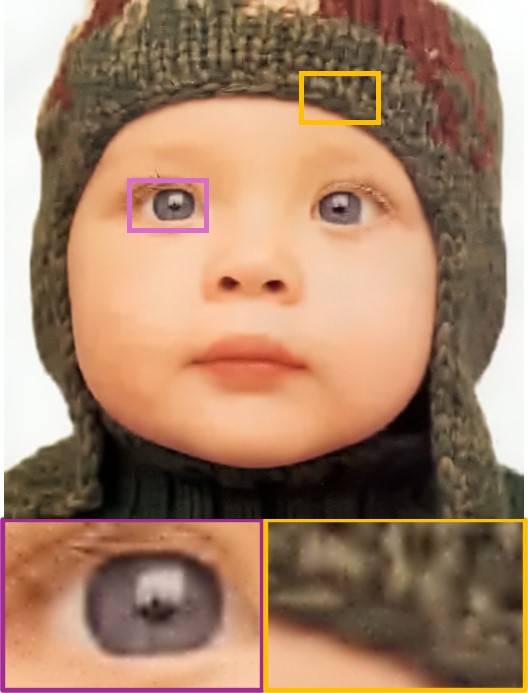} \hfill & 
         \includegraphics[width=2.8cm] 
        {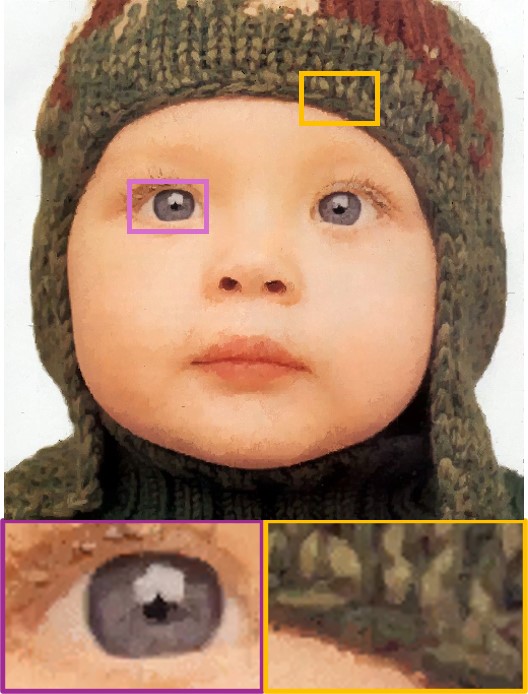} \hfill & 
        \includegraphics[width=2.8cm] 
        {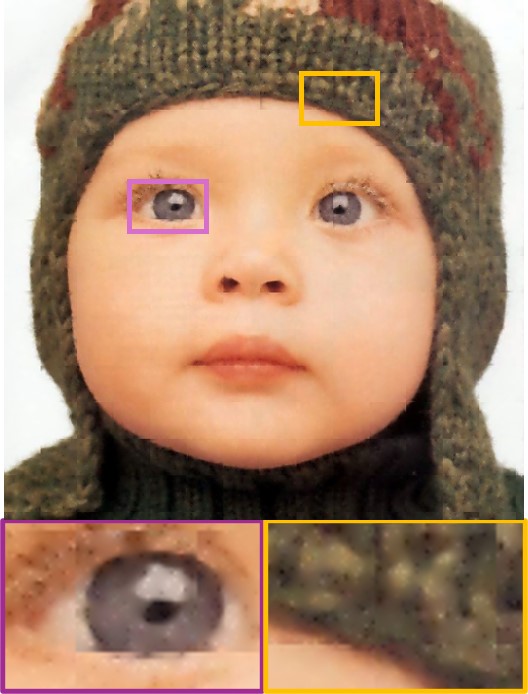} \hfill & 
        \includegraphics[width=2.8cm] 
        {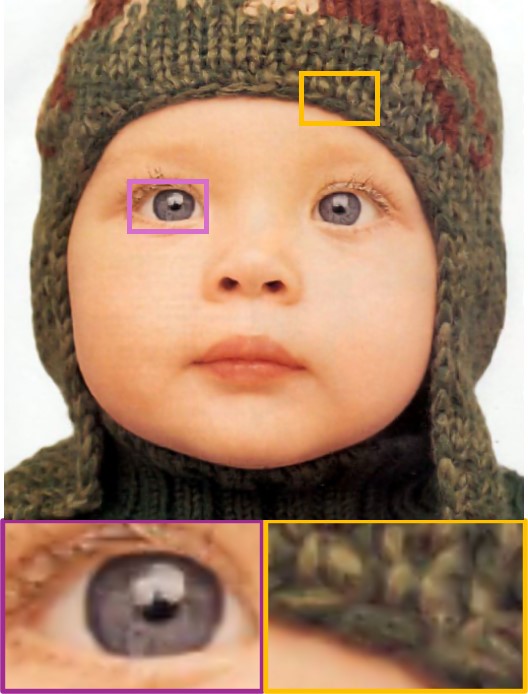} \hfill & 
        \includegraphics[width=2.8cm] 
        {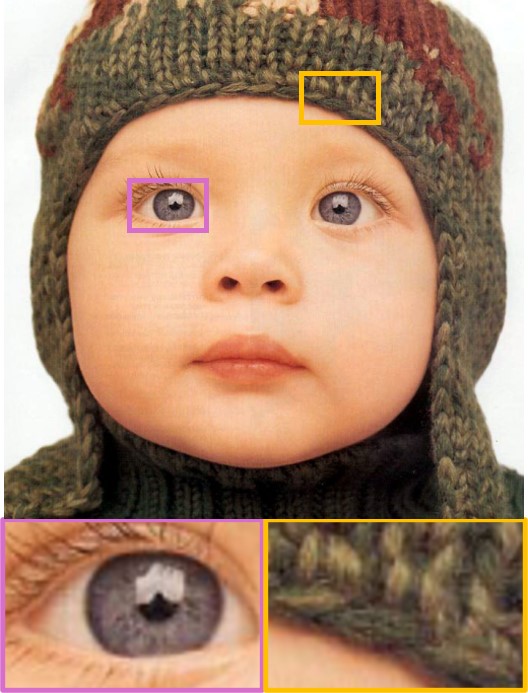} \\

       Masked &  EPLL / 31.07dB  &  IRCNN / 31.13dB & DeepGLR / 30.86dB & UPnPGGLR / 32.48dB & Ground-truth\\

          \includegraphics[width=2.8cm] 
        {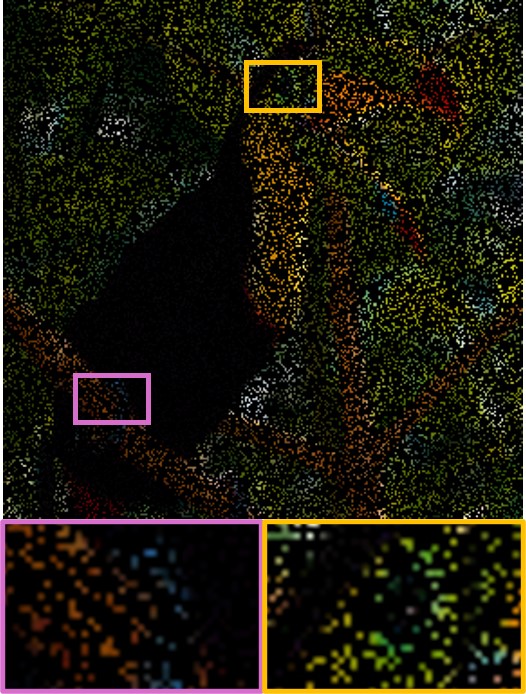} \hfill & 
         \includegraphics[width=2.8cm] 
        {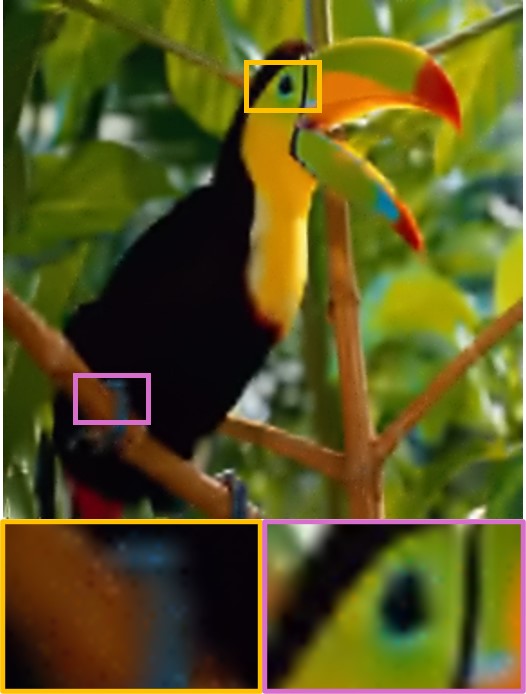} \hfill & 
         \includegraphics[width=2.8cm] 
        {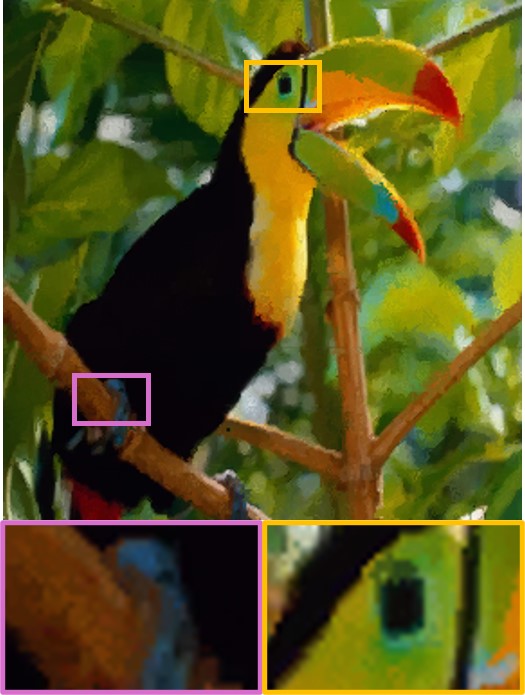} \hfill & 
        \includegraphics[width=2.8cm] 
        {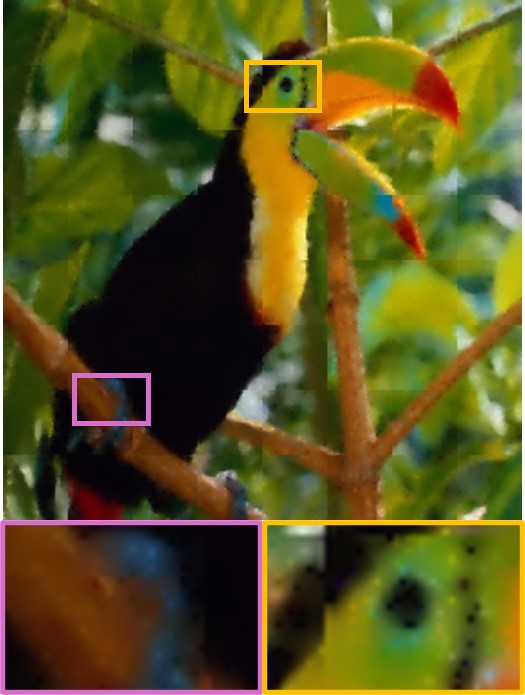} \hfill & 
        \includegraphics[width=2.8cm] 
        {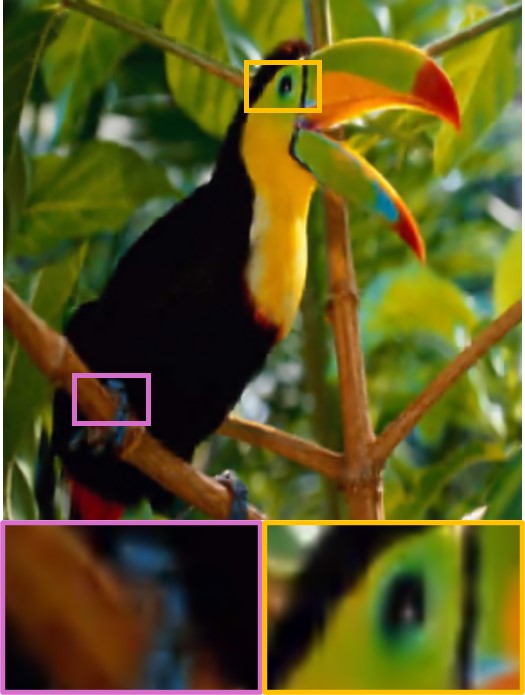} \hfill & 
        \includegraphics[width=2.8cm] 
        {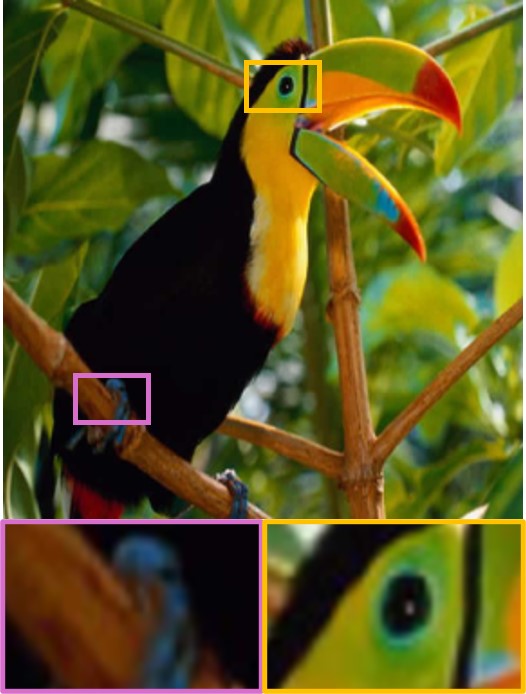} \\

       Masked &  EPLL / 29.39dB  &  IRCNN / 29.89dB & DeepGLR / 28.63dB & UPnPGGLR / 31.41dB & Ground-truth\\

	\end{tabular}
	\caption{Visual comparision for interpolating $80\%$ of missing pixels on the Set5 dataset \cite{timofte2015a+}. Interpolation by UPnPGGLR looks less blocky and more natural.}
	\label{fig:visual_interpolation}
\end{figure*}

%% file: image/curve/curve.tex
\begin{figure*}[t]
\footnotesize

\begin{center}

\begin{tabular}{c@{}c@{}c@{}c@{}c@{}c}

  \includegraphics[width=5.8cm] 
        {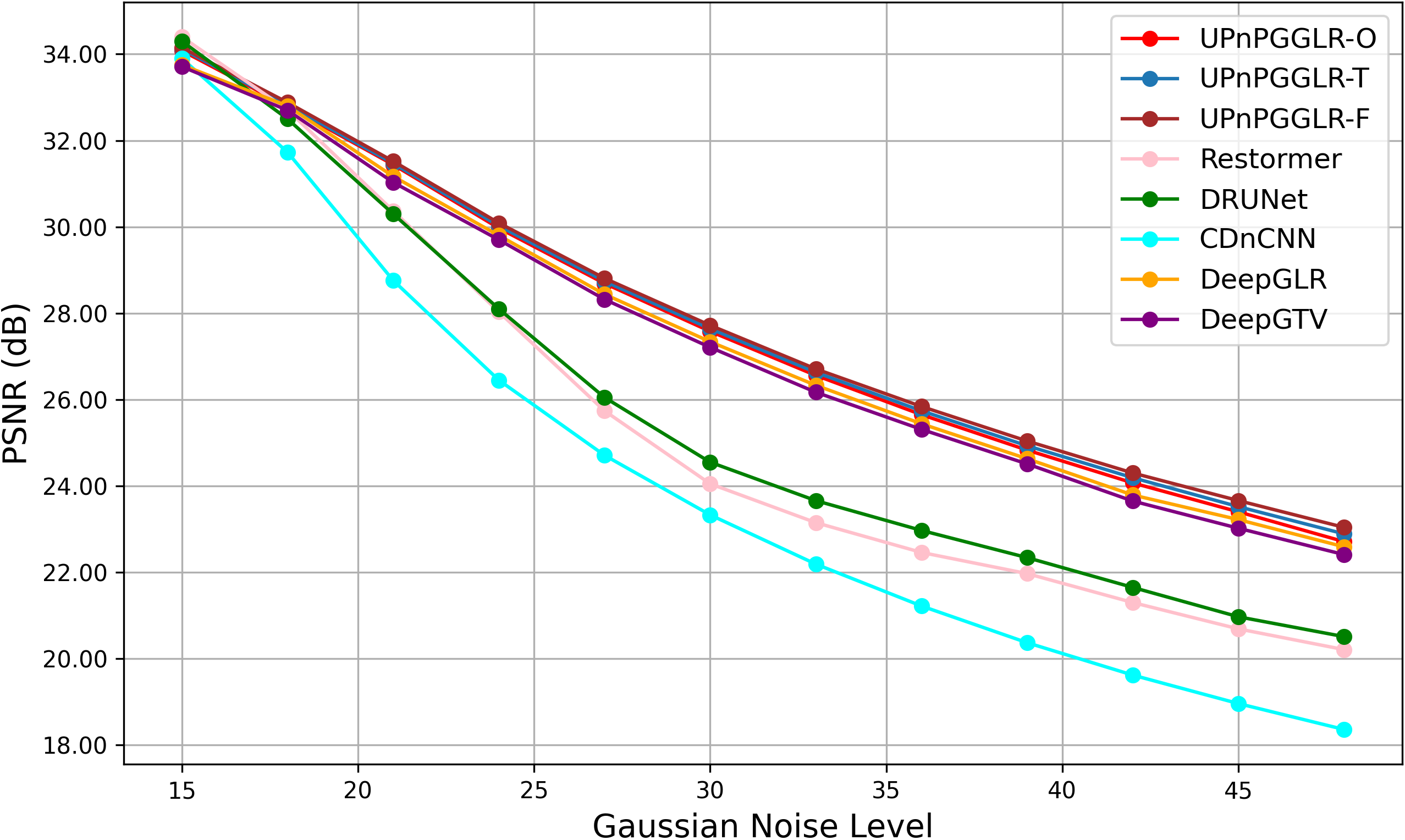} \hspace{-0.04mm} & 
		  \includegraphics[width=5.8cm] 
        {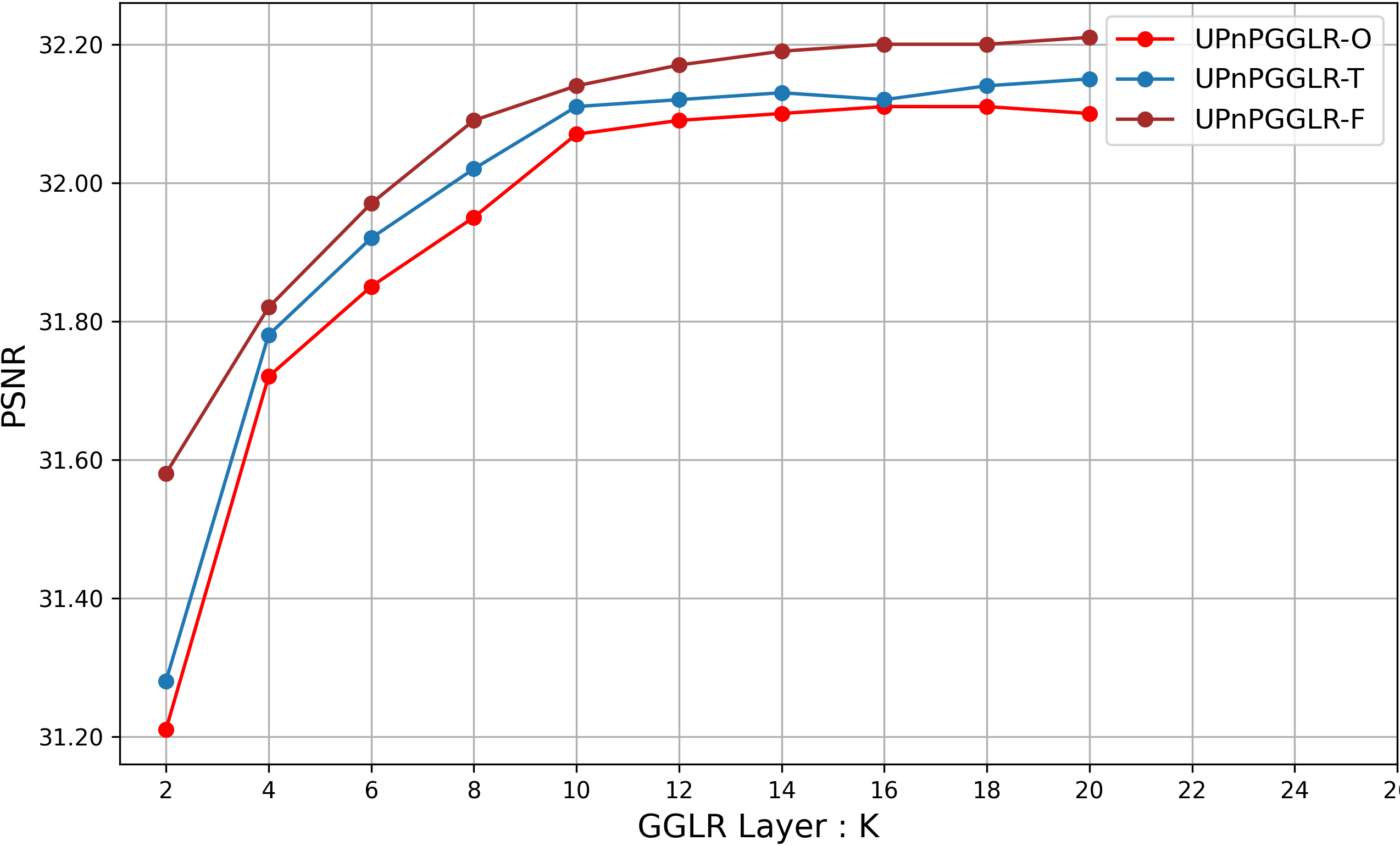} \hspace{-0.04mm} & 
		  \includegraphics[width=5.8cm] 
        {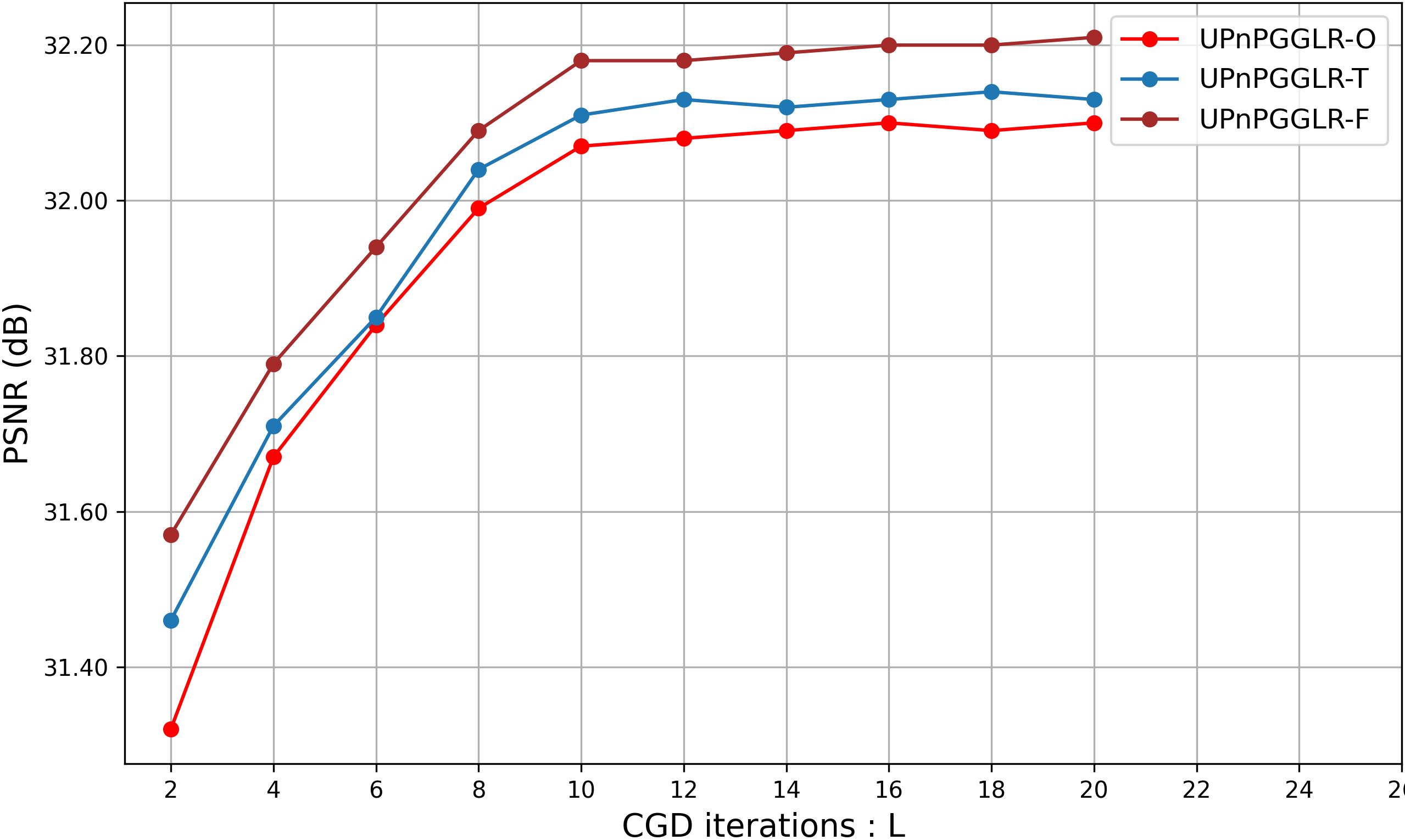} \hspace{-0.04mm}  \\

         (a) Covariance shift & (b) GGLR Layer : $K$  & (c) CGD Iteration : $L$
\end{tabular}
\end{center}

\caption{\blue{Image quality in PSNR versus the number of GGLR layers and CGD iterations on RENOIR dataset.}}
\vspace{-0.05in}
\label{fig:curve}
\end{figure*}

 

 %



%% file: table/table_flops.tex
\begin{table*}[h]
\caption{\blue{Parameters (in M), Runtime (in seconds) and FLOPs (in G) of different methods on images of size 480×320 with noise level 50.}}
\vspace{-0.2in}
\label{table_flop}
\begin{center}
\setlength{\tabcolsep}{3pt}
\resizebox{0.8\linewidth}{!}{
\begin{tabular}{c|ccccccc}
\hline
Metric       & CDnCNN \cite{DnCNN}  & IRCNN \cite{zhang2017learning} & DeepGLR \cite{deepglr} & DeepGTV \cite{deepgtv} & DRUNet \cite{zhang2021plug} & Restormer \cite{Zamir2021RestormerET}  & UPnPGGLR\\ \hline
Paras(M)    & 0.56   & 4.75  & 0.93    & \textbf{0.10}    & 32.64  & 26.11      & 0.23    \\
Runtime(Sec) & \textbf{0.13}   & 0.47  & 9.01    & 7.86    & 0.27   & 0.28       & 9.02   \\
FLOPs(G)     & 102.99 & 28.99 & 8.71    & \textbf{4.08}    & 336.58 & 344.35    & 10.43 \\ \hline
\end{tabular}
}
\vspace{-0.1in}
\end{center}
\end{table*}

%% file: table/interpolation_table.tex


\begin{table*}[ht]
\vspace{-0.05in}
\caption{Image interpolation results (PSNR/SSIM) by different methods on Set5 \cite{timofte2015a+}.}
\vspace{-0.2in}
\label{table_interpolation}

\begin{center}
\setlength{\tabcolsep}{3pt}
\resizebox{0.8\linewidth}{!}{
\begin{tabular}{c|ccccccc}
\hline
Method & EPLL \cite{zoran2011learning}       & IRCNN \cite{zhang2017learning}      & DeepGLR \cite{deepglr}     & DeepGTV \cite{deepgtv} & \blue{SNORE \cite{SNORE}}   & \blue{UPnPGGLR-S} & UPnPGGLR     \\ \hline
Paras(M)                 & 1.65   & 4.75  & 0.93  &  0.10  & \blue{32.64} & \blue{\textbf{0.09}} & 0.23 \\ \hline
20\% missing               & 37.23/0.974 & 39.20/0.981 & 38.92/0.981 & 38.85/0.980 & \textbf{\blue{40.01/0.985}} & \blue{39.32/0.981} & 39.88/0.984 \\ 
50\% missing               & 33.25/0.943 & 33.84/0.949 & 33.75/0.948 & 33.65/0.948 &  \blue{34.66/ 0.951} & \blue{34.37/0.950} & \textbf{34.81/0.952} \\
80\% missing               & 27.87/0.833 & 28.32/0.864 & 27.48/0.843 & 27.42/0.842 & \blue{21.96/0.736} & \blue{28.48/0.867} & \textbf{28.98/0.877} \\ \hline
\end{tabular}
}
\vspace{-0.3in}
\end{center}
\end{table*}

%% file: image/deblurry/deblur/deblur_more.tex
\begin{figure*}[t]
    \centering

    \vspace{0.15in}
     \fontsize{6}{10}\selectfont
	\begin{tabular}{c@{}c@{}c@{}c@{}c@{}c@{}c@{}}



           \includegraphics[width=2.8cm] 
        {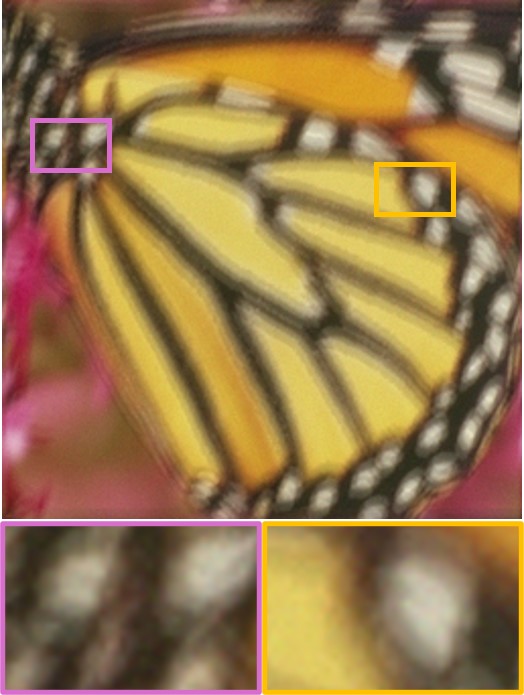} \hfill & 
         \includegraphics[width=2.8cm] 
        {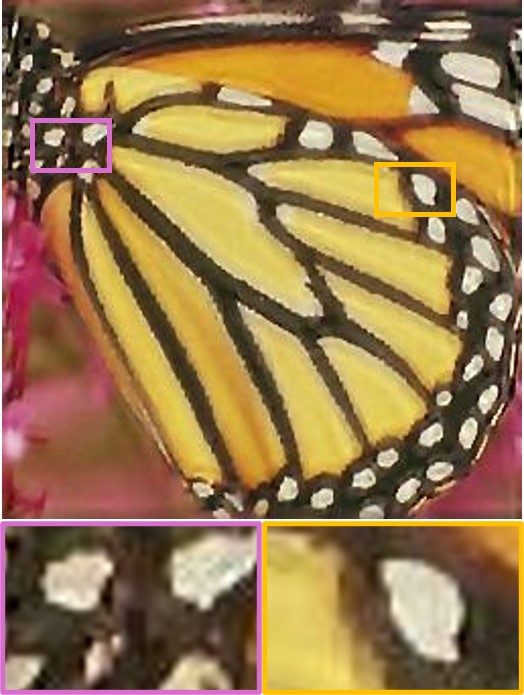} \hfill & 
         \includegraphics[width=2.8cm] 
        {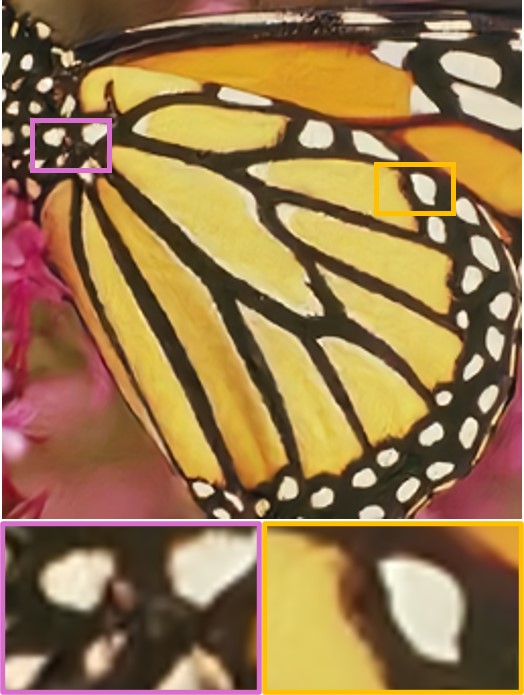} \hfill & 
        \includegraphics[width=2.8cm] 
        {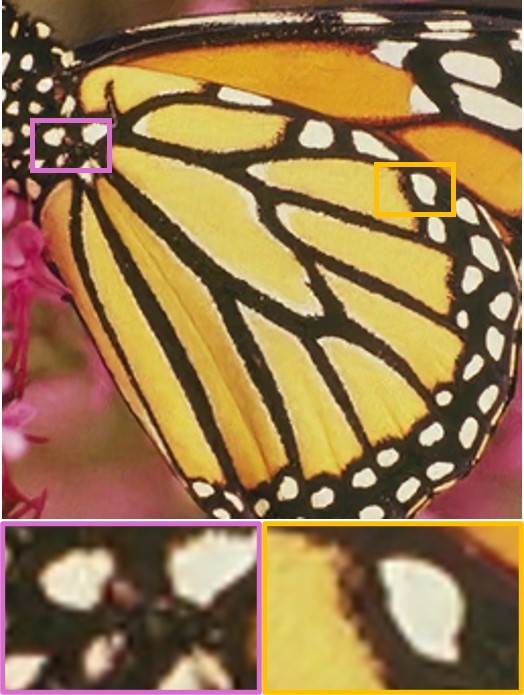} \hfill & 
        \includegraphics[width=2.8cm] 
        {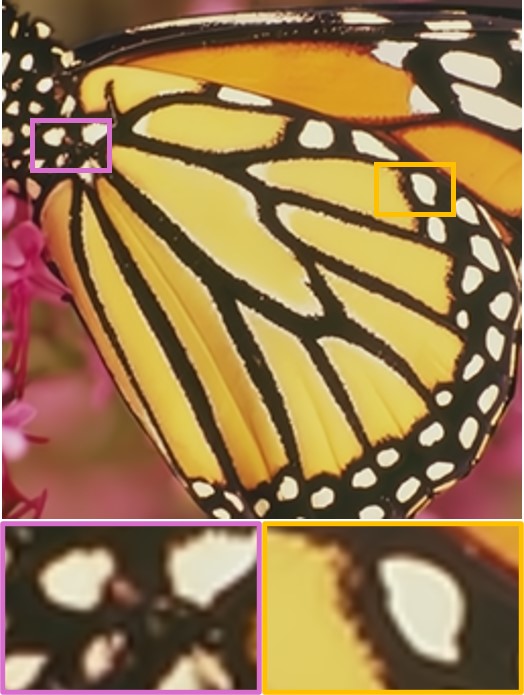} \hfill & 
        \includegraphics[width=2.8cm] 
        {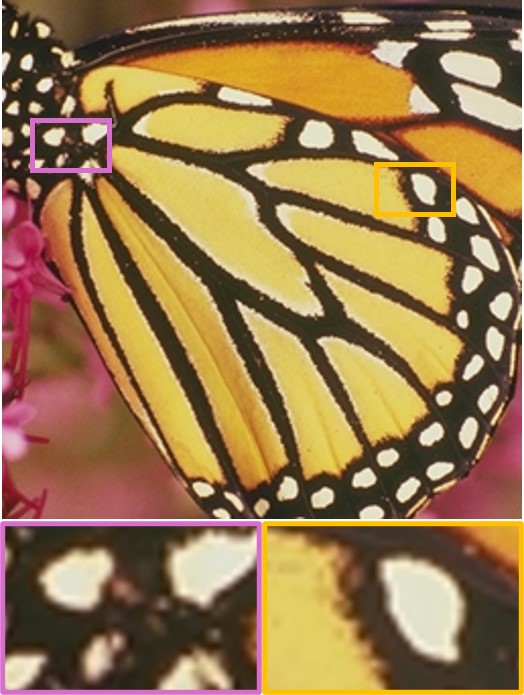} \\

        Blurry&  RGTV / 28.78dB  &  FDN / 28.99dB & IRCNN / 33.11dB & UPnPGGLR / 34.88dB & Ground-truth \\



           \includegraphics[width=2.8cm] 
        {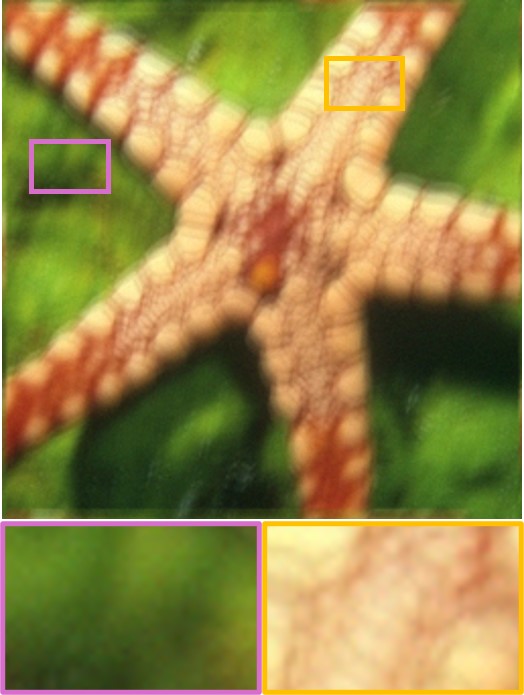} \hfill & 
         \includegraphics[width=2.8cm] 
        {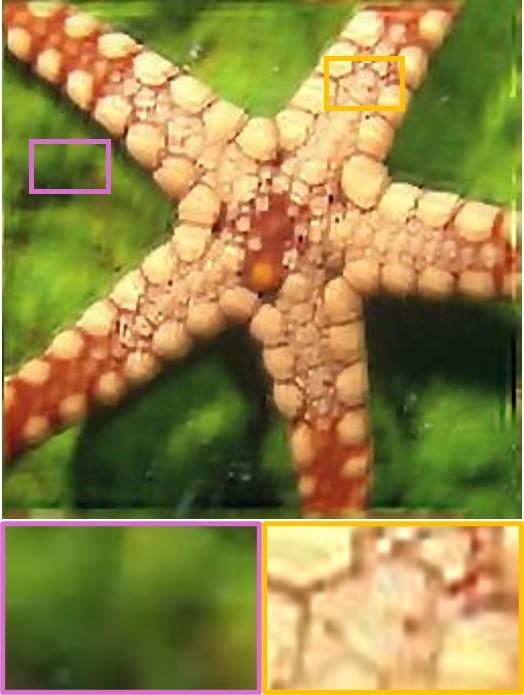} \hfill & 
         \includegraphics[width=2.8cm] 
        {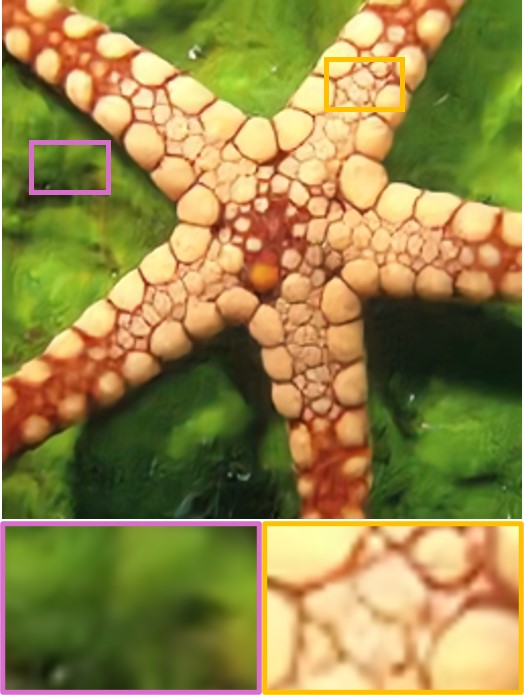} \hfill & 
        \includegraphics[width=2.8cm] 
        {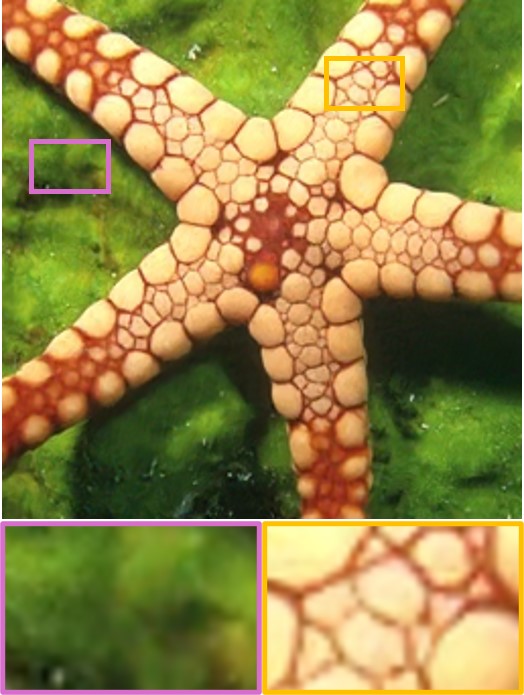} \hfill & 
        \includegraphics[width=2.8cm] 
        {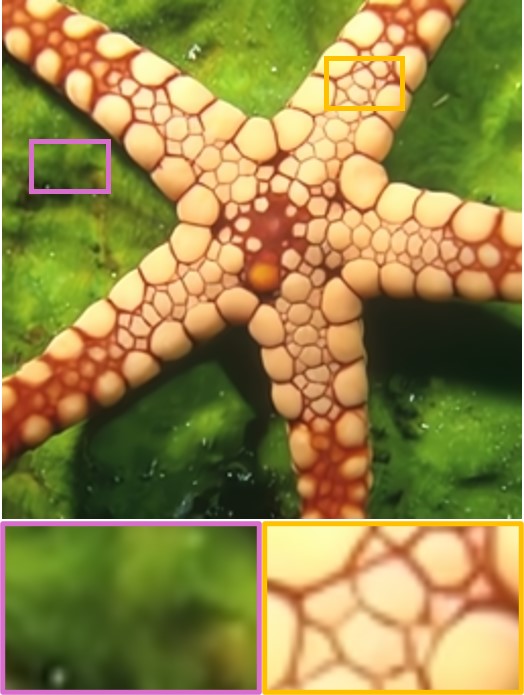} \hfill & 
        \includegraphics[width=2.8cm] 
        {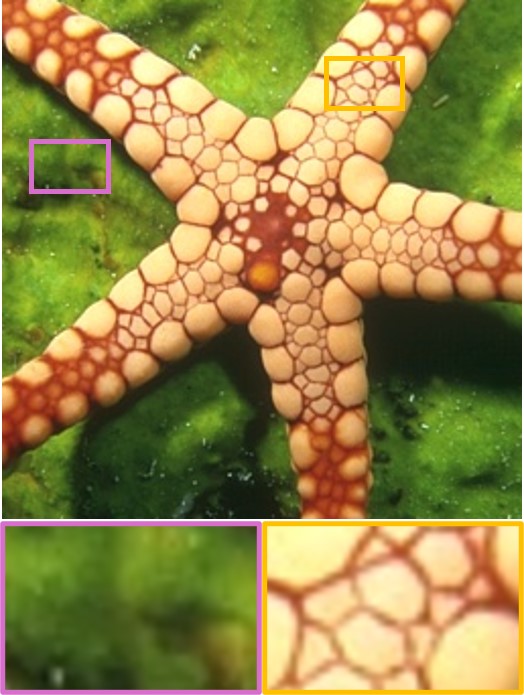} \\

        Blurry &  RGTV / 28.30dB  & FDN / 28.66dB & IRCNN / 32.82dB & UPnPGGLR / 34.49dB & Ground-truth\\


	\end{tabular}
	\caption{Visual results comparison of different deblurring methods on Set6 dataset \cite{zhang2017learning}. The blur kernel size is $19\times19$ and the noise level is 2.55 ($1\%$).}
	\label{fig:visual_deblur}
\end{figure*}

%% file: table/deblurry_table.tex

\begin{table*}[t]
\caption{PSNR/SSIM results of different methods on Set6 \cite{zhang2017learning} for image deblurring.}
\vspace{-0.2in}
\label{table_deblur}
\begin{center}
\setlength{\tabcolsep}{3pt}
\resizebox{0.85\linewidth}{!}{
\begin{tabular}{c|cccccccc}
\hline
Method         & EPLL \cite{zoran2011learning}        & RGTV \cite{bai2018blind}       & FDN \cite{kruse2017learning}         & IRCNN \cite{zhang2017learning} & \blue{Eq.DRUNet \cite{EqDRUNtet}} & \blue{SNORE \cite{SNORE}}    & \blue{UPnPGGLR-S}  & UPnPGGLR    \\ \hline
Paras(M)  & 1.65 &  - & 0.39 &4.75 &\blue{32.64} & \blue{32.64}& \blue{\textbf{0.09}} & 0.23  \\  \hline
Kernel 1 19×19 & 27.88/0.843 & 28.54/0.849 & 28.94/0.856 & 33.02/0.922 & \blue{29.16/0.863} & \blue{31.98/0.907} & \blue{34.07/0.933}  & \textbf{34.57/0.941} \\
Kernel 2 17×17 & 27.56/0.840 & 27.98/0.845 & 28.53/0.851 & 32.43/0.911 & \blue{29.23/0.865} & \blue{31.76/0.904} & \blue{33.98/0.932}  & \textbf{34.57/0.941} \\ \hline
\end{tabular}
}
\vspace{-0.1in}
\end{center}
\end{table*}

%% file: image/parameter.tex
\begin{figure*}[h]
\footnotesize
\centering
\begin{tabular}{c@{}c@{}c@{}c@{}c@{}c}
        \includegraphics[width=5.4cm] 
        {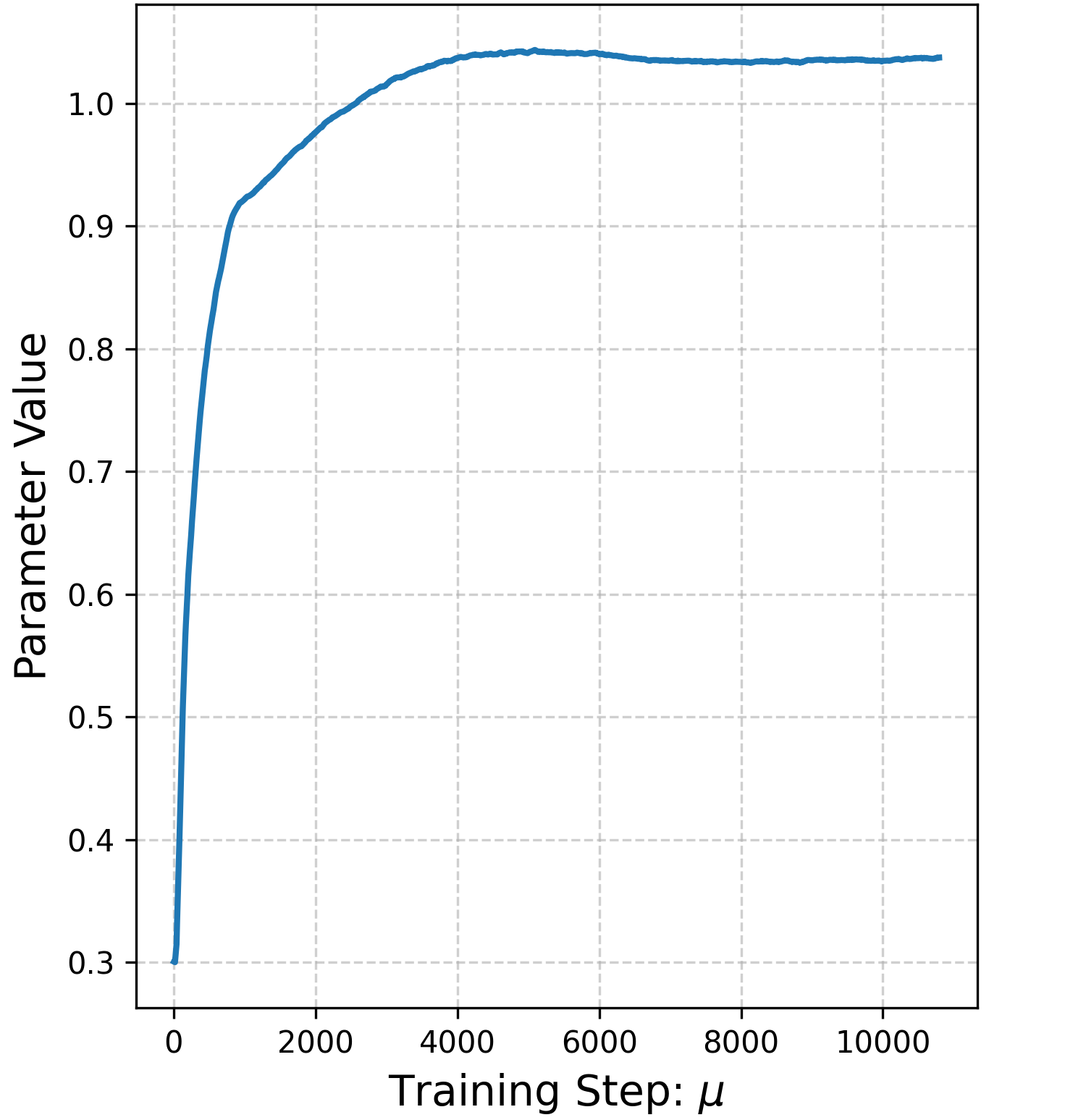} \hspace{-0.04mm} &
		  \includegraphics[width=5.4cm] 
        {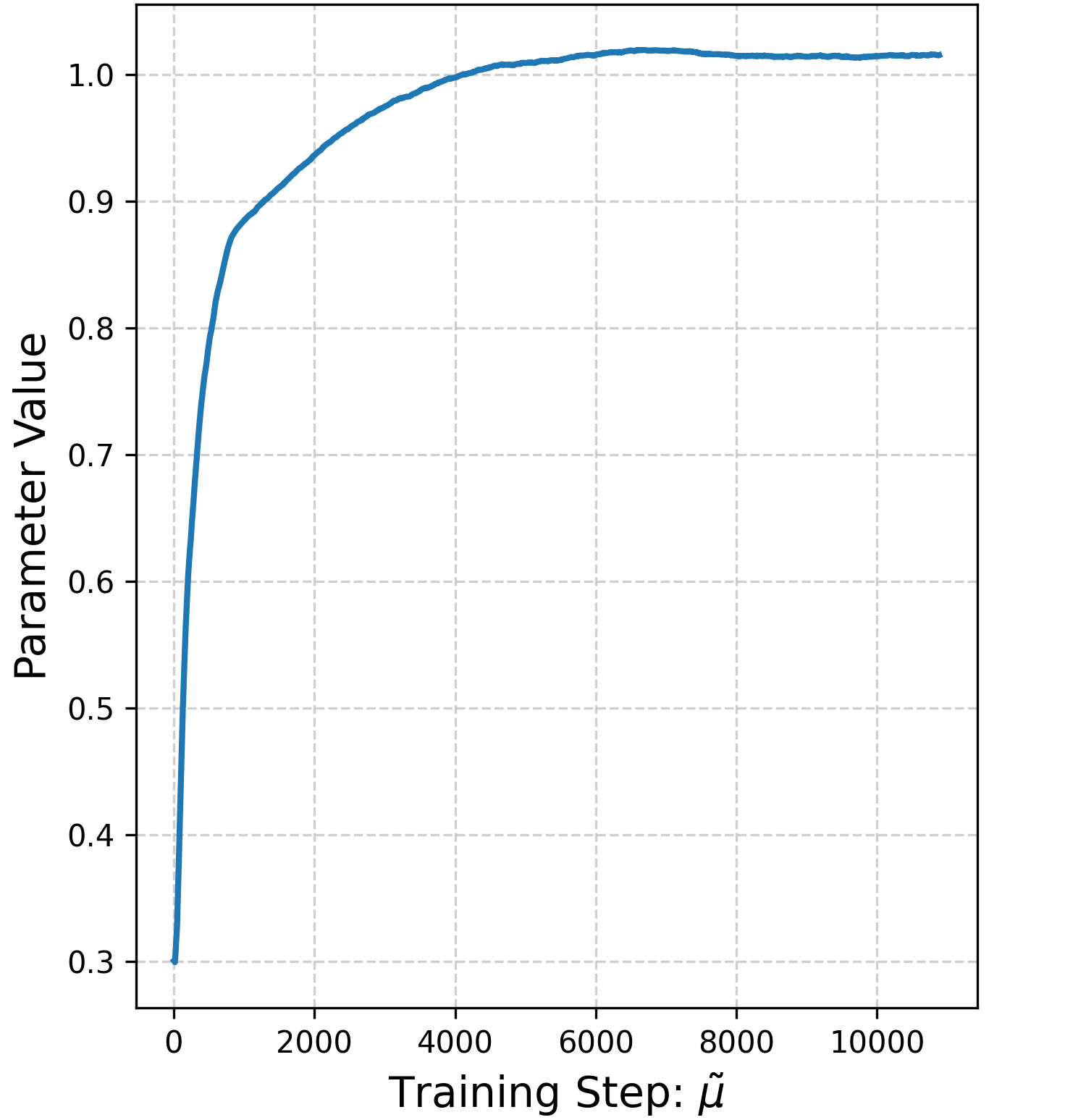} \hspace{-0.04mm} & 
		  \includegraphics[width=5.4cm] 
        {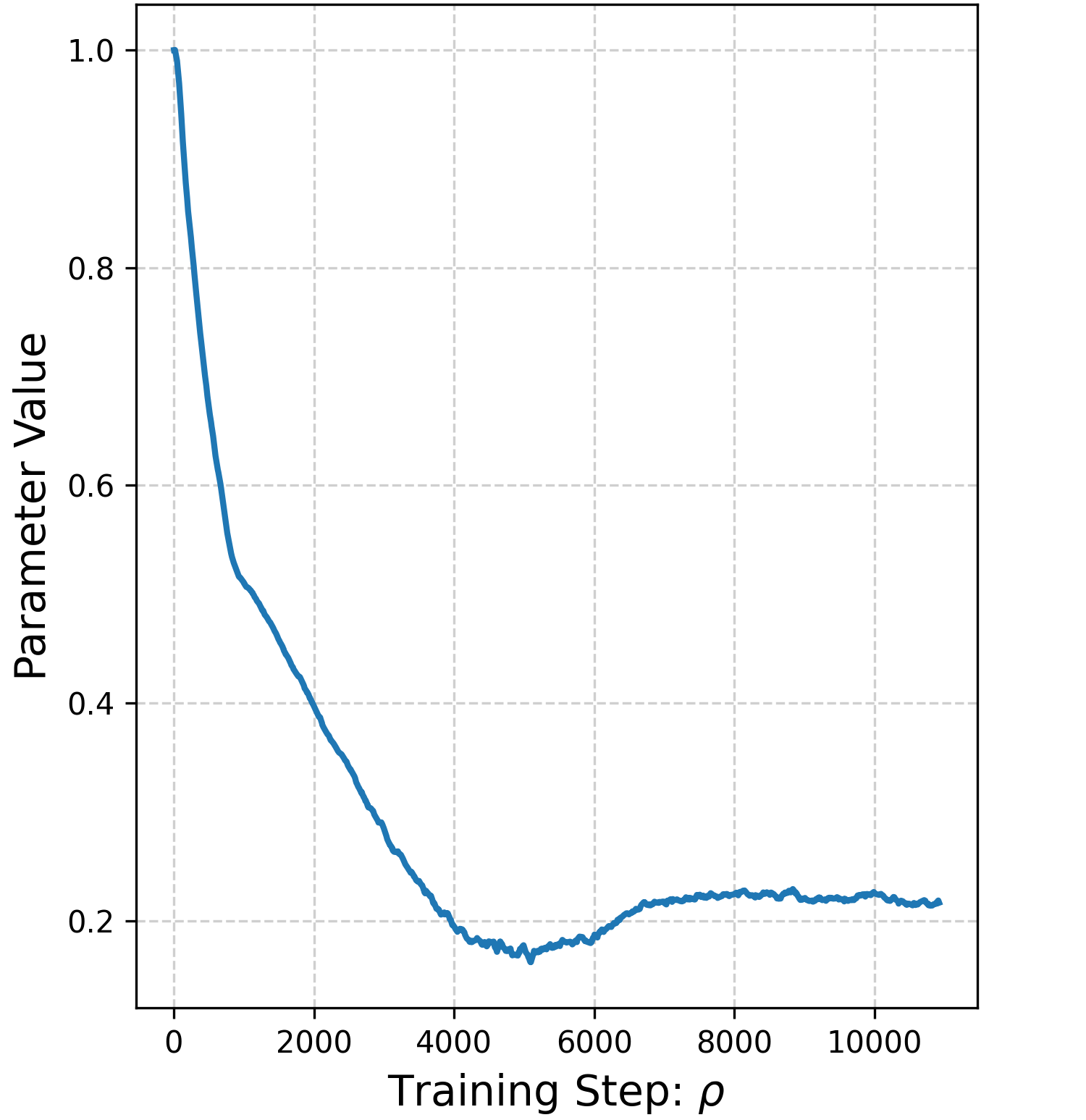} \hspace{-0.04mm}  \\
          (a) Iteration : $\mu$ &  (b) Iteration : $\tilde{\mu}$  & (c) Iteration : $\rho$
\end{tabular}
\caption{\blue{Plots for ablation experiments showing the evolution of the learnable parameters $\mu$, $\tilde{\mu}$ and $\rho$ across iterations.}}

\vspace{-0.05in}
\label{fig:pamameter}
\end{figure*}

%% file: conclude.tex
Towards interpretable and lightweight deep neural nets for image restoration, we design a family of graph-based ADMM optimization algorithms by varying the number of introduced auxiliary variables and unroll them into feed-forward networks for data-driven parameter tuning.
This is possible thanks to a recent graph smoothness prior GGLR (promoting PWP reconstruction) that is a sum of priors applied separately to individual pixel rows and columns of a target patch. 
We argue that our periodically inserted graph learning module is akin to the self-attention mechanism in a conventional transformer architecture, but by adopting shallow CNNs to learn low-dimensional feature representations, our variant is dramatically more parameter-efficient. 
Experimental results show that our unrolled networks perform competitively to SOTA in various image restoration tasks, while using only a small fraction of parameters. 
Moreover, our model demonstrates improved robustness to covariate shift.

\blue{Our proposed UPnPGGLR has some limitations. First, our current framework is limited to image formation models that are linear in the form $\y = \A \x + \n$, where $\n$ is an additive Gaussian noise.
Formation models that are non-linear, or with noise that are not additive Gaussian, are currently not directly applicable.
Second, the inference time of our method is relatively slow. 
This is due to the current lack of efficient implementations of sparse matrix-vector multiplication.
Future work will focus on addressing these limitations.}

%% file: append.tex
\subsection{GLR Promoting PWC Signal Reconstruction}
\label{append:GLR}

Different from arguments in \cite{pang17,liu17}, we argue here that iterative GLR promotes \textit{piecewise constant} (PWC) signal reconstruction from an anisotropic diffusion perspective. 
When denoising noise-corrupted observation $\y \in \mathbb{R}^N$, solution $\x^* \in \mathbb{R}^N$ is the argument minimizing the following objective regularized using the \textit{graph Laplacian regularizer} (GLR) \cite{pang17} $\x^\top \L \x$ as signal prior given graph Laplacian matrix $\L \in \mathbb{R}^{N \times N}$ corresponding to a positive graph $\cG$:
\begin{align}
\x^* = \arg \min_\x \|\y - \x\|^2_2 + \mu \, \x^\top \L \x 
\end{align}
where $\mu > 0$ is a weight parameter trading off the fidelity term and the signal prior.
$\x^*$ can be computed via the inverse of a positive definite (PD) matrix $\I_N + \mu \L$, given $\mu > 0$ and $\L$ is a provably positive semi-definite (PSD) Laplacian matrix to a positive graph $\cG$ \cite{cheung18}:
\begin{align}
\x^* = (\I_N + \mu \L)^{-1} \y \stackrel{(a)}{\approx} \sum_{m=0}^M \frac{a_m}{s^{m+1}} (\I_N + \mu \L - s \I_N)^m \y
\end{align}
where $(a)$ is a truncated \textit{Taylor Series Expansion} (TSE) of the matrix inverse function of $M+1$ terms, $a_m = (-1)^m$ are the TSE coefficients, and $s$ is the fixed point at which the approximation is computed.
For simplicity, consider the asymmetric random-walk Laplacian $\L_{rw} \triangleq \D^{-1} \L$ instead of $\L$ ($\D$ is a diagonal degree matrix), with real eigenvalues\footnote{Random-walk Laplacian $\L_{rw} = \D^{-1} \L$ is a similarity transform of symmetric normalized Laplacian $\L_n = \D^{-1/2} \L \D^{-1/2}$ that has real eigenvalues in range $[0,2]$ \cite{ortega18ieee}.} $\lambda_i \in [0, 2]$. 
Thus, fixed point $s$ can be chosen as the midpoint of the eigenvalue range $[1, 1+ 2\mu]$ for $\I_N + \mu \L_{rw}$, \ie, $s=1+\mu$. 
Setting $M=1$ we get
\begin{align}
\x^* &\approx \left(\frac{1}{1+\mu} \I_N + \frac{-1}{(1+\mu)^2} (\mu \L_{rw} - \mu \I_N) \right) \y 
\\
&= \left( \frac{\I_N + \mu \I_N - \mu \L_{rw} + \mu \I_N}{(1+\mu)^2} \right) \y \\
&= (1+\mu)^{-2} \left( (1+2\mu) \I_N - \mu \L_{rw} \right) \y .
\label{eq:TSE_approx}
\end{align}
Given that $\L_{rw}$ is a high-pass filter, \eqref{eq:TSE_approx} states that output $\x^*$ is a scaled version of $\y$ filtered by low-pass filter $(1+2\mu)\I_N - \mu \L_{rw}$. 

Suppose now that the underlying graph $\cG$ is a line graph, and hence $\L_{rw}$ is a tri-diagonal matrix. 
Then,
\begin{align}
x_i^* &= (1+\mu)^{-2} \left[ (1+2\mu) y_i 
\right. \nonumber \\
& ~~~~~~~~~~~~~~~~ \left. -\mu (-w_{i-1,i} y_{i-1} + y_{i} - w_{i,i+1} y_{i+1} ) \right] \\
&= (1+\mu)^{-2} \left[ (1+2\mu) y_i 
\right. \nonumber \\
& ~~~~~~~~~~~ \left. - \mu \left( w_{i-1,i}(y_i - y_{i-1}) - w_{i,i+1} (y_{i+1} - y_i) \right) \right]
\end{align}  
where $w_{i-1,i} + w_{i,i+1} = 1$ due to normalization in $\L_{rw}$. 
Defining \textit{discrete gradient} $\nabla y_i \triangleq y_i - y_{i-1}$, 
we rewrite $x_i^*$ as
\begin{align}
x_i^* &= (1+\mu)^{-2} \left( (1+2\mu) y_i - \mu \left( w_{i-1,i} \nabla y_i - w_{i,i+1} \nabla y_{i+1} \right) \right) .
\end{align}

If this graph filter is applied iteratively (with edge weights recomputed based on updated signal in each iteration), then at iteration $t$, 

\vspace{-0.05in}
\begin{small}
\begin{align}
x_{i}^{t+1} \!\! - x_i^t &= (1+\mu)^{-2} \left[ (1+2\mu) x_i^t 
\right. \nonumber \\
& ~~~~~~~~~~~~ \left. - \mu \left( w_{i-1,i} \nabla x_{i}^t - w_{i,i+1} \nabla x_{i+1}^t \right) \right] - x_i^t \\
&= (1+\mu)^{-2} \left( (- \mu^2) x_i^t  + \mu \left( w_{i-1,i} \nabla x_{i}^t - w_{i,i+1} \nabla x_{i+1}^t \right) \right)  \\
&\stackrel{(a)}{\approx} \frac{\mu}{(1+\mu)^{2}} \left( w_{i-1,i} \nabla x_{i}^t - w_{i,i+1} \nabla x_{i+1}^t \right)
\label{eq:Laplacian_diffusion}
\end{align}
\end{small}\noindent
where in $(a)$ we ignore the first term scaled by $\mu^2$, which is a good approximation when $\mu$ is small.
Recall that edge weight $w_{i,i+1}$ is defined in a signal-dependent manner:

\vspace{-0.05in}
\begin{small}
\begin{align}
w_{i,i+1} &= \exp \left( - \frac{\|\f_i - \f_{i+1}\|^2_2}{\sigma_f^2} - \frac{|x_i - x_{i+1}|}{\sigma_x^2} \right) 
\nonumber \\
&= \exp \left( - \frac{\|\f_i - \f_{i+1}\|^2_2}{\sigma_f^2} - \frac{|\nabla x_{i+1}|}{\sigma_x^2} \right) 
\end{align}
\end{small}
\noindent
where $w_{i,i+1} \rightarrow 0$ when $|\nabla x_{i+1}| \rightarrow \infty$.
Similarly done in \cite{barash02}, we compare \eqref{eq:Laplacian_diffusion} to anisotropic diffusion proposed by Perona and Malik \cite{perona90}:
\begin{align}
\frac{\partial x}{\partial t} &= c(x) \nabla x
\label{eq:PM_diffusion}
\end{align}
where $c(x)$ is the nonlinear diffusion coefficient, typically defined as
\begin{align}
c(x) = g(\|\nabla x\|)
\label{eq:edge_preserve}
\end{align}
where $g(\cdot)$ is an \textit{edge preserving function}, \ie, $g(s) \rightarrow 0$ when $s \rightarrow \infty$. 
By interpreting signal-dependent edge weights $w_{i,i+1}$ as nonlinear diffusion coefficients that is edge preserving \eqref{eq:edge_preserve}, we see now that \eqref{eq:Laplacian_diffusion} is a variant of anisotropic diffusion similar to \eqref{eq:PM_diffusion}. 
Given that Perona and Malik's anisotropic diffusion (\textit{total varation} (TV) \cite{chambolle97} is an example) is known to promote PWC signal reconstruction---$\frac{\partial x}{\partial t} = 0$ in \eqref{eq:PM_diffusion} if $\nabla x = 0$ (constant signal) or if $c(x) = g(\|\nabla x\|) = 0$ (signal discontinuity), iterative GLR with signal-dependent edge weights also promotes PWC signal reconstruction.

\subsection{GGLR Promoting PWL Signal Reconstruction}
\label{append:GGLR}

Given that iterative GLR with signal-dependent edge weights promotes PWC signal reconstruction, as discussed in Appendix\;\ref{append:GLR}, applying iterative GLR to horizontal/vertical gradient of a pixel row/column would mean PWC gradient reconstruction.
A piecewise linear (PWL) signal has PWC gradient, and thus promotion of PWC gradient reconstruction via GGLR implies promotion of PWL signal reconstruction.


\subsection{PnP ADMM for Two Auxiliary Variables}
\label{append:twoAux}

Recall that the ADMM objective function when two auxiliary variables $\z$ and $\tilde{\z}$ are introduced is
\begin{align}
\min_{\x,\z,\tilde{\z}}  & ~\|\y - \A \x\|^2_2 + \mu \,  \z^\top \cL \z + \blambda^\top (\x - \z) + \frac{\rho}{2} \left\|\x - \z \right\|^2_2  
\nonumber \\
& + \tilde{\mu} \,  \tilde{\z}^\top \tilde{\cL} \tilde{\z} + \tilde{\blambda}^\top (\x - \tilde{\z}) + \frac{\tilde{\rho}}{2} \left\|\x - \tilde{\z} \right\|^2_2 .
\label{eq:obj_ADMM_a2_append}
\end{align}
We solve \eqref{eq:obj_ADMM_a2_append} iteratively by solving for variables $\x$, $\z$ and $\tilde{\z}$ and updating Lagrange multipliers $\tilde{\blambda}$ and $\tilde{\blambda}$ in order as
\begin{align}
\x^{(k+1)} &= \arg \min_\x ~ \left\| \y -\A \x \right\|_{2}^{2} + \frac{\rho}{2} \| \x-\hat{\x}^{(k)} \|^2_2 
\label{eq:ADMM1_append} \\
\z^{(k+1)} &= \arg \min_\z ~ \mu \, \z^\top \cL \z + \frac{\rho}{2} \|\z - \hat{\z}^{(k)} \|^2_2
\label{eq:ADMM2_append} \\
\tilde{\z}^{(k+1)} &= \arg \min_{\tilde{\z}} ~ \mu \, \tilde{\z}^\top \tilde{\cL} \tilde{\z} + \frac{
\tilde{\rho}}{2} \|\tilde{\z} - \hat{\tilde{\z}}^{(k)} \|^2_2
\label{eq:ADMM3_append} \\
\hat{\blambda}^{(k+1)} &= \hat{\blambda}^{(k)} + \left(\x^{(k+1)} - \z^{(k+1)} \right)
\label{eq:ADMM4_append} \\
\hat{\tilde{\blambda}}^{(k+1)} &= \hat{\tilde{\blambda}}^{(k)} + \left(\x^{(k+1)} - \tilde{\z}^{(k+1)} \right)
\label{eq:ADMM4_append}
\end{align}
where 
\begin{align}
\hat{\blambda}^{(k)} &= (1/\rho) \blambda^{(k)} 
\\
\hat{\x}^{(k)} &= \z^{(k)} - \hat{\blambda}^{(k)} + \tilde{\z}^{(k)} - \hat{\tilde{\blambda}}^{(k)}  \\ 
\hat{\tilde{\blambda}}^{(k)} &= (1/\rho) \tilde{\blambda}^{(k)} 
\\
\hat{\z}^{(k)} &= \x^{(k+1)} + \hat{\blambda}^{(k)} 
\\
\hat{\tilde{\z}}^{(k)} &= \x^{(k+1)} + \hat{\tilde{\blambda}}^{(k)}.
\end{align}
\eqref{eq:ADMM1_append}, \eqref{eq:ADMM2_append} and \eqref{eq:ADMM3_append} are convex quadratic equations whose respective solutions $\x^{(k+1)}$, $\z^{(k+1)}$ and $\tilde{\z}^{(k+1)}$ can be obtained by solving linear systems: 
\begin{align}
(2\A^\top \A + \rho \I_N) \x^{(k+1)} &= 2\A^\top \y + \rho \, \hat{\x}^{(k)} 
\label{eq:linSys1_append} \\
(\I_N + \frac{2 \mu}{\rho} \cL) \z^{(k+1)} &= \hat{\z}^{(k)} 
\label{eq:linSys2_append} \\
(\I_N + \frac{2 \tilde{\mu}}{\tilde{\rho}} \tilde{\cL}) \tilde{\z}^{(k+1)} &= \hat{\tilde{\z}}^{(k)} .
\label{eq:linSys3_append}
\end{align}
These linear systems can be solved using \textit{conjugate gradient} (CG) \cite{conjugate_grad} efficiently without matrix inverse, as described in Section\;\ref{subsubsec:LG}.